\documentclass[a4paper,11pt]{article}
\pdfoutput=1
\usepackage{jcappub}
\usepackage[T1]{fontenc}
\usepackage{ifthen}
\usepackage{graphicx}
\usepackage{amsmath,amssymb}
\usepackage{MnSymbol}

\usepackage[nolist]{acronym}
\newacro{AR}[AR]{anti-reflective}
\newacro{ARC}[ARC]{anti-reflective coating}
\newacro{CES}[CES]{constant-elevation scan}
\newacro{CMB}[CMB]{cosmic microwave background}
\newacro{CRHWP}[CRHWP]{continuously rotating half-wave plate}
\newacro{DAQ}[DAQ]{data acquisition}
\newacro{FIR}[FIR]{finite impulse response}
\newacro{HTT}[HTT]{Huan Tran telescope}
\newacro{HWP}[HWP]{half-wave plate}
\newacro{HWPSS}[HWPSS]{\ac{HWP} synchronous signal}
\newacro{I2P}[$I{\rightarrow}P$]{intensity to polarization}
\newacro{PCA}[PCA]{principal component analysis}
\newacro{PSD}[PSD]{power-spectrum density}
\newacroplural{PSD}[PSDs]{power-spectrum densities}
\newacro{PWV}[PWV]{precipitable water vapor}
\newacro{TES}[TES]{transition edge sensor}
\newacro{TOD}[TOD]{time-ordered data}

\ifthenelse{\isundefined{\arcmin}}{
\newcommand{\arcmin}{${}'$}}{}
\newcommand{\pb}{\textsc{Polarbear}\ }

\ifthenelse{\isundefined{\aap}}{
\newcommand\aap{A\&A}

}{}

\title{Performance of a continuously rotating half-wave plate on the POLARBEAR telescope}

\author[1,2]{Satoru Takakura,}
\author[3]{Mario Aguilar,}
\author[4,2]{Yoshiki Akiba,}
\author[5]{Kam Arnold,}
\author[6,7]{Carlo Baccigalupi,}
\author[8]{Darcy Barron,}
\author[8]{Shawn Beckman,}
\author[9]{David Boettger,}
\author[10,11]{Julian Borrill,}
\author[12]{Scott Chapman,}
\author[8,13]{Yuji Chinone,}
\author[8]{Ari Cukierman,}
\author[13]{Anne Ducout,}
\author[5]{Tucker Elleflot,}
\author[14,15,16]{Josquin Errard,}
\author[17,6,7]{Giulio Fabbian,}
\author[18,13]{Takuro Fujino,}
\author[5]{Nicholas Galitzki,}
\author[8]{Neil Goeckner-Wald,}
\author[19,20,21]{Nils W. Halverson,}
\author[2,4]{Masaya Hasegawa,}
\author[22,2]{Kaori Hattori,}
\author[2,4,13,23]{Masashi Hazumi,}
\author[8,24]{Charles Hill,}
\author[5]{Logan Howe,}
\author[25,2]{Yuki Inoue,}
\author[26]{Andrew H. Jaffe,}
\author[8]{Oliver Jeong,}
\author[13]{Daisuke Kaneko,}
\author[13]{Nobuhiko Katayama,}
\author[5]{Brian Keating,}
\author[10,11]{Reijo Keskitalo,}
\author[10,11]{Theodore Kisner,}
\author[6]{Nicoletta Krachmalnicoff,}
\author[24]{Akito Kusaka,}
\author[8,24]{Adrian T. Lee,}
\author[5]{David Leon,}
\author[5]{Lindsay Lowry,}
\author[5]{Frederick Matsuda,}
\author[13]{Tomotake Matsumura,}
\author[5]{Martin Navaroli,}
\author[2]{Haruki Nishino,}
\author[5]{Hans Paar,}
\author[27]{Julien Peloton,}
\author[6]{Davide Poletti,}
\author[6,7]{Giuseppe Puglisi,}
\author[28]{Christian L. Reichardt,}
\author[12]{Colin Ross,}
\author[5]{Praween Siritanasak,}
\author[8,29]{Aritoki Suzuki,}
\author[2,4]{Osamu Tajima,}
\author[4,2]{Sayuri Takatori,}
\author[5]{and Grant Teply}

\affiliation[1]{Department of Earth and Space Science, Graduate School of Science, Osaka University, Toyonaka, Osaka, 560-0043 Japan}
\affiliation[2]{High Energy Accelerator Research Organization (KEK), Tsukuba, Ibaraki, 305-0801 Japan}
\affiliation[3]{Departamento de F\'isica, FCFM, Universidad de Chile, Blanco Encalada 2008, Santiago, Chile}
\affiliation[4]{SOKENDAI (The Graduate University for Advanced Studies), Hayama, Miura District, Kanagawa, 240-0115 Japan}
\affiliation[5]{Department of Physics, University of California, San Diego, CA, 92093-0424 U.S.A.}
\affiliation[6]{International School for Advanced Studies (SISSA), Via Bonomea 265, Trieste, I-34136 Italy}
\affiliation[7]{INFN, Sezione di Trieste, Via Valerio 2, Trieste, I-34127 Italy}
\affiliation[8]{Department of Physics, University of California, Berkeley, CA, 94720 U.S.A.}
\affiliation[9]{Centro de Astro-Ingenier\'ia, Pontificia Universidad Cat\'olica de Chile, Vicu\~na Mackenna 4860, Santiago, Chile}
\affiliation[10]{Computational Cosmology Center, Lawrence Berkeley National Laboratory, Berkeley, CA, 94720 U.S.A.}
\affiliation[11]{Space Sciences Laboratory, University of California, Berkeley, CA, 94720 U.S.A.}
\affiliation[12]{Department of Physics and Atmospheric Science, Dalhousie University, Halifax, NS, B3H 4R2 Canada}
\affiliation[13]{Kavli IPMU (WPI), UTIAS, The University of Tokyo, Kashiwa, Chiba, 277-8583 Japan}
\affiliation[14]{Sorbonne Universit\'es, Institut Lagrange de Paris (ILP), 98 bis Boulevard Arago, Paris, 75014 France}
\affiliation[15]{LPNHE, CNRS-IN2P3 and Universit\'es Paris 6 \& 7, 4 place Jussieu, Paris Cedex 05, F-75252 France}
\affiliation[16]{AstroParticule et Cosmologie, Universit\'e Paris Diderot, CNRS/IN2P3, CEA/Irfu, Obs de Paris, Sorbonne, Paris Cit\'e, France}
\affiliation[17]{Institut d'Astrophysique Spatiale, CNRS (UMR 8617), Universit\'{e} Paris-Sud, Universit\'e Paris-Saclay, b\^{a}t. 121, Orsay, 91405 France}
\affiliation[18]{Yokohama National University, Yokohama, Kanagawa, 240-8501 Japan}
\affiliation[19]{Center for Astrophysics and Space Astronomy, University of Colorado, Boulder, CO, 80309 U.S.A.}
\affiliation[20]{Department of Astrophysical and Planetary Sciences, University of Colorado, Boulder, CO, 80309 U.S.A.}
\affiliation[21]{Department of Physics, University of Colorado, Boulder, CO, 80309 U.S.A.}
\affiliation[22]{National Institute of Advanced Industrial Science and Technology (AIST), Tsukuba, Ibaraki, 305-8563 Japan}
\affiliation[23]{Institute of Space and Astronautical Science (ISAS), Japan Aerospace Exploration Agency (JAXA), Sagamihara, Kanagawa, 252-0222 Japan}
\affiliation[24]{Physics Division, Lawrence Berkeley National Laboratory, Berkeley, CA, 94720 U.S.A.}
\affiliation[25]{Institute of Physics, Academia Sinica, 128, Sec.~2, Academia Road, Nankang, Taiwan}
\affiliation[26]{Department of Physics, Blackett Laboratory, Imperial College London, London, SW7 2AZ U.K.}
\affiliation[27]{Department of Physics \& Astronomy, University of Sussex, Brighton, BN1 9QH U.K.}
\affiliation[28]{School of Physics, University of Melbourne, Parkville, Victoria, 3010 Australia}
\affiliation[29]{Radio Astronomy Laboratory, University of California, Berkeley, CA, 94720 U.S.A.}

\emailAdd{takakura@vega.ess.sci.osaka-u.ac.jp}

\abstract{
A continuously rotating half-wave plate (CRHWP) is a promising tool to improve the sensitivity to large angular scales in cosmic microwave background (CMB) polarization measurements. 
With a CRHWP,
single detectors can measure three of the Stokes parameters,
$I$, $Q$ and $U$, 
thereby avoiding the set of systematic errors that can be introduced by mismatches in the properties of orthogonal detector pairs.
We focus on the implementation of CRHWPs in large aperture telescopes (i.e. the primary mirror is larger than the current maximum half-wave plate diameter of $\sim0.5\,\mathrm{m}$),
where the CRHWP can be placed between the primary mirror and focal plane.
In this configuration, one needs to address the intensity to polarization ($I{\rightarrow}P$) leakage of the optics, which becomes a source of 1/f noise and also causes differential gain systematics that arise from CMB temperature fluctuations.
In this paper, we present the performance of a CRHWP installed in the \textsc{Polarbear} experiment, which employs a Gregorian telescope with a $2.5\,\mathrm{m}$ primary illumination pattern.
The CRHWP is placed near the prime focus between the primary and secondary mirrors. 
We find that the $I{\rightarrow}P$ leakage is larger than the expectation from the physical properties of our primary mirror, resulting in a 1/f knee of $100\,\mathrm{mHz}$.
The excess leakage could be due to imperfections in the detector system, 
i.e. detector non-linearity in the responsivity and time-constant.
We demonstrate, however, that by subtracting the leakage correlated with the intensity signal, 
the 1/f noise knee frequency is reduced to $32\,\mathrm{mHz}$ $(\ell \sim 39$ for our scan strategy), which is very promising to probe the primordial B-mode signal. 
We also discuss methods for further noise subtraction in future projects where the precise temperature control of instrumental components and the leakage reduction will play a key role.
}

\keywords{CMBR experiments -- gravitational waves and CMBR polarization}

\toccontinuoustrue
\begin{document}
\maketitle
\section{Introduction}
\label{sec:introduction}
Degree-scale \acf{CMB} B-mode polarization from primordial gravitational waves is a unique probe of the beginning of the Universe~\citep{SeljakZaldarriaga1997PhRvL}.
B-modes can test the validity of cosmic inflationary models and the underlying quantum gravity theories~\citep{Lyth1997PhRvL}.
Sub-degree scale CMB polarization measurements are also sensitive to gravitational lensing~\citep{ZaldarriagaSeljak1998PhRvD}. 
Precision measurements of these lensing B-modes will allow us to determine the sum of neutrino masses~\citep{Neutrino2015APh}. By observing degree-scale and sub-degree-scale B-modes simultaneously, we can subtract the lensing B-modes and improve the sensitivity to the primordial B-modes~\citep{HuOkamoto2002ApJ}.
In order to achieve these goals, foreground signals need to be subtracted, 
for which observations with multiple frequencies are necessary~\citep{PlanckXXX2016A&A}.

In current published results (see e.g.~\cite{ACTPol2016Cl}),
degree-scale observations and sub-degree scale observations are independently performed by small-aperture experiments and large-aperture experiments, respectively.
One impediment is the presence of low-frequency noise, also called 1/f noise, that degrades the sensitivity of large aperture experiments to degree-scale B-modes.
Large-aperture experiments have the potential to observe degree-scale B-modes if the detector 1/f noise is low enough.
Therefore, a demonstration of improved 1/f noise control with large aperture experiments is desired.
Future experiments will contain tens or even hundreds of thousands of detectors and improve the sensitivity by an order of magnitude.
Further reduction of the 1/f noise will be required to achieve the full potential of these projects.

Polarization modulation using a
\acf{CRHWP} is a well-known technique to reduce the impact of both 1/f noise and instrumental systematic errors.\footnote{Other techniques to modulate polarization signal can be found in e.g. refs.~\citep{CMBPol2009JPhCS,Pisano2014SPIE}.}
Rotating the polarization angle to which the detectors are sensitive,
we can measure the modulated polarization signal in a frequency band, where the detector sensitivity is dominated by white noise that is limited by the photon noise. 
Additionally, the CRHWP enables a detector, which is sensitive to a single linear polarization state,
to measure both $Q$ and $U$ Stokes parameters.
Observations using a CRHWP are expected to have less systematic uncertainties than methods that take the difference between orthogonal detectors which may have different properties~\cite[e.g.][]{ABS2016Systematics}.
Therefore, 
the CRHWP has the potential to be one of the essential tools for the next-generation CMB experiments, such as CMB-S4~\citep{CMBS4},
a next-generation ground-based experiment,
and LiteBIRD, 
a proposed space mission for full-sky CMB polarization measurements~\citep{LiteBIRD2016SPIE}. 

Many previous studies on \acp{HWP}
have been carried out on millimeter and submillimeter experiments,
including MAXIPOL~\citep{MAXIPOL2007HWP}, EBEX~\citep{EBEX2010}, SPIDER~\citep{SPIDER2008SPIE}, BLASTPol~\citep{BLASTPolHWP}, ABS~\citep{ABS2014Demod}, \textsc{Polarbear}~\citep{POLARBEAR2012SPIE} and Advanced ACTPol~\citep{AdvACTPol2015LTD}.
Many types of HWPs have been developed and their performance measured in laboratories~\cite[e.g.][]{Hanany2005,Pisano2006,Savini2006,Pisano2008,Savini2009,Matsumura2009,SPIDER2010HWP,BLASTPolHWP}. 
In particular,
Refs.~\cite{Pisano2003,O'Dea2007,Brown2009,SPIDER2010Mueller,O'Dea2011,Essinger-Hileman2013,ABS2016Systematics}
have modeled the non-ideality of an actual HWP.

One of the most important concerns in implementing a CRHWP is 
\acf{I2P} leakage 
in the optics, which could cause 1/f noise and systematic errors~\cite{ABS2014Demod,ABS2016Systematics}. 
Since the \ac{I2P} leakage is caused by asymmetry in the optical elements on the sky side of the CRHWP,
it is ideally placed as far skyward as possible in the telescope's optical system,
as in the case of ABS~\citep{ABS2014Demod}, to prevent the \ac{I2P} leakage. 
If we aim to measure both B-mode signals from the primordial gravitational waves and weak gravitational lensing, however, the required angular resolution is $\sim0.1^\circ$ or better. 
At $150\,\mathrm{GHz}$, 
which is one of the optimal frequencies for CMB observations from the ground,
the required aperture size for this angular resolution is too large to locate a CRHWP on the sky-side of the primary aperture,
as the maximum diameter of HWP available now is $\sim 50\,\mathrm{cm}$~\citep{Hill2016SPIE}.

MAXIPOL~\citep{MAXIPOL2007HWP}, a balloon-borne experiment, 
had a $1.3\,\mathrm{m}$ primary mirror and put a CRHWP after three mirrors. They reported their pioneering measurements of the \ac{I2P} leakage. 
The sensitivity of current state-of-the-art focal plane arrays is,
however, better than that of MAXIPOL by an order of magnitude, 
and ground-based experiments suffer from larger atmospheric 1/f noise than balloon-borne experiments. 
We thus need to deal with much more stringent requirements on \ac{I2P} leakage.

\pb is 
a ground-based telescope in operation in Atacama, Chile. It has a $2.5\,\mathrm{m}$ aperture and employs a CRHWP between the primary and secondary mirrors.
Science observations with the CRHWP started in May 2014.
This paper reports on the performance of a prototype CRHWP during a test observation in February 2014.
We used the data collected to examine the \ac{I2P} leakage due to the primary mirror and non-linearities of the detector.
We evaluate the \ac{I2P} leakage in three ways: from the amplitude of the polarized emission from optical elements, so-called instrumental polarization (method A), from the dependence of the instrumental polarization on the total intensity of the atmosphere (method B), and from the correlation between the intensity and polarization timestreams (method C).
We also demonstrate the leakage subtraction using our own intensity timestream as a template, which sufficiently mitigates the impact of the leakage on the 1/f noise and the systematic bias in the B-mode angular power spectrum for the current observation with \textsc{Polarbear}.

In this article, we first review the basics of the CRHWP in section~\ref{sec:CRHWP}.
In section~\ref{sec:measurements}, 
we describe our measurements of the \ac{I2P} leakage and 1/f noise at \textsc{Polarbear}.
In section~\ref{sec:discussion}, we discuss the origin of the leakage as well as methods to further mitigate the leakage and 1/f noise in future projects. 
We summarize our studies in section~\ref{sec:summary}.

\section{CRHWP and intensity to polarization leakage}\label{sec:CRHWP}
The \ac{CRHWP}
enables a single polarization-sensitive detector to measure Stokes $I$, $Q$ and $U$ polarization and reduces the impact of 1/f noise and systematics originating from the differential properties between orthogonal detectors.

In practice,
however, the optical elements on the sky side of the CRHWP and non-idealities of the HWP can create the \ac{I2P} leakage,
which becomes the source of both 1/f noise~\cite[see, e.g.][]{ABS2014Demod} and systematic errors~\citep{MAXIPOL2007HWP,ABS2016Systematics}.
Additionally, 
we find that the non-linearity of the detector couples with the finite amplitude of the instrumental polarization and mimics
\ac{I2P} leakage.
In the following sections, we focus on the \ac{I2P} leakage and the instrumental polarization, and how to measure them.

\subsection{Polarization modulation}\label{sec:modulation}
In this section, we briefly review the basic principles of CRHWP operation.
See \cite{ABS2014Demod} for more details.

A \ac{HWP} is an optical element that rotates the polarization angle of incident light.
If the incident radiation has a polarization angle $\theta_\mathrm{in}$,
the polarization angle of the outgoing radiation becomes $2\theta_\mathrm{HWP}-\theta_\mathrm{in}$ after passing through an ideal HWP with birefringent axis angle $\theta_\mathrm{HWP}$.
By rotating the HWP at an angular velocity of $\omega_\mathrm{HWP}$, we can modulate the polarization angle by $2\theta_\mathrm{HWP}(t)=2\omega_\mathrm{HWP}t$.
We can express the effect in terms of Stokes parameters as
\begin{equation}
Q_\mathrm{out}(t) - {i} U_\mathrm{out}(t) = [Q_\mathrm{in}(t) + {i} U_\mathrm{in}(t)] \, m(t)\:,
\end{equation}
where $m(t) \equiv e^{-{i}\omega_\mathrm{mod}t}$ is the modulation function at the modulation frequency $\omega_\mathrm{mod}\equiv4\omega_\mathrm{HWP}$.
Here, the additional factor of two comes from the spin-2 nature of linear polarization.
$Q_\mathrm{out}(t)$, $U_\mathrm{out}(t)$, $Q_\mathrm{in}(t)$ and $U_\mathrm{in}(t)$ are Stokes $Q$ and $U$ of the incident and outgoing radiations, respectively.
The Stokes $I$ and $V$ are not modulated in this ideal HWP case.
The HWP-modulated signal $d_m(t)$ measured by a polarization-sensitive detector becomes
\begin{equation}\label{eq:d_m}
d_m(t) = \delta I_\mathrm{in}(t) + \varepsilon \mathrm{Re}\{[Q_\mathrm{in}(t) + {i} U_\mathrm{in}(t)] \, m(t) \, e^{-2{i}\theta_\mathrm{det}} \} + \mathcal{N}_{m}(t)\:,
\end{equation}
where $\varepsilon$ is a polarization modulation efficiency, $\theta_\mathrm{det}$ is an angle of the detector, and $\mathcal{N}_{m}(t)$ is the white noise of a detector. 
Note that only the fluctuation of the total intensity, $\delta I_\mathrm{in}(t) \equiv I_\mathrm{in}(t)-\langle I_\mathrm{in}\rangle $, where the angle bracket denotes the average of the data, is well characterized.
Since there is a large uncertainty in the absolute value of the total intensity,\footnote{ Large uncertainties in the total intensity are primarily due to drifts in the focal plane temperature and variable loading from the cryogenic environment.
} the baseline is usually filtered out.
If the rotation frequency is faster than the scan velocity divided by the beam size,\footnote{ This condition is satisfied in our configuration (see \autoref{sec:observation}).
Without this condition, we need to combine data from different detectors to solve the mixing of Stokes parameters (see~\cite{O'Dea2007}).
} we can adopt a demodulation method in which the intensity and polarization signals are extracted from the corresponding frequency bands of the modulated signal as
\begin{align}
&d_{\bar{m}}(t) \equiv \mathcal{F}_\mathrm{LPF}\{d_m(t)\} = \delta I_\mathrm{in}(t) + \mathcal{N}_{\bar{m}}(t) \:,\label{eq:d0f}\\
&d_{\bar{d}}(t) \equiv \mathcal{F}_\mathrm{LPF}\{d_m(t) \, 2m^*(t) \, e^{2 {i} \theta_\mathrm{det}} \} = \varepsilon [ Q_\mathrm{in}(t) + {i} U_\mathrm{in}(t) ]+ \mathcal{N}_{\bar{d}}^{(\mathrm{Re})}(t) + {i} \mathcal{N}_{\bar{d}}^{(\mathrm{Im})}(t)\:.
\label{eq:demodmethod}\end{align}
Here, $d_{\bar{m}}(t)$ and $d_{\bar{d}}(t)$ are the measured signals of intensity and polarization, 
respectively, and $m^*(t)$ is the complex-conjugate of $m(t)$.
$\mathcal{F}_\mathrm{LPF}$ is a low-pass filter, whose cutoff frequency is typically the rotation frequency or its double.
The white noise $\mathcal{N}_{m}(t)$ is split into three white noises, $\mathcal{N}_{\bar{m}}(t)$, $\mathcal{N}_{\bar{d}}^{(\mathrm{Re})}(t)$ and $\mathcal{N}_{\bar{d}}^{(\mathrm{Im})}(t)$, where the variance of latter two is double of the former as $\langle [\mathcal{N}_{\bar{d}}^{(\mathrm{Re})}(t)]^2 \rangle =\langle [\mathcal{N}_{\bar{d}}^{(\mathrm{Im})}(t)]^2 \rangle = 2 \langle [\mathcal{N}_{\bar{m}}(t)]^2 \rangle \equiv2N$. 

\subsection{Instrumental polarization and \ac{I2P} leakage}\label{sec:IPandleakage}
Instrumental polarization\footnote{Instrumental polarization includes both the polarized emission from the optical elements and the polarized signal due to the \ac{I2P} leakage.
We only consider the effect from the optical elements before the CRHWP.
The instrumental polarization from the optical elements after the CRHWP is less important, since it is not modulated at first order.}
and non-idealities of the HWP\footnote{Non-idealities of the HWP include polarized emission along the birefringent axis, \ac{I2P} leakage due to differential transmission between the birefringent axes, non-uniformity of the 
\ac{ARC},
eccentricity of the rotator, non-perpendicular incident angle, etc. } create 
\acp{HWPSS}.
The HWPSSs may have stable contributions from the instrument and leakage components proportional to the fluctuation of incident intensity.
Thus, the HWPSSs, $A(\theta_\mathrm{HWP} ,t)$ may be modeled as
\begin{equation}\label{eq:HWPSS_definition}
A(\theta_\mathrm{HWP} ,t) \equiv \sum_{n=1} \frac{1}{2}\left[ A_{0|\langle I_\mathrm{in}\rangle }^{(n)}+\lambda^{(n)}_\mathrm{opt} \delta I_\mathrm{in}(t)\right]e^{-{i} n \theta_\mathrm{HWP}} + \frac{1}{2}\left[ A_{0|\langle I_\mathrm{in}\rangle }^{(n)*}+\lambda^{(n)*}_\mathrm{opt} \delta I_\mathrm{in}(t)\right]e^{{i} n \theta_\mathrm{HWP}}\:.
\end{equation}
Here we have decomposed the signal into harmonics of the HWP angle. 
$A_{0|\langle I_\mathrm{in}\rangle }^{(n)}$ is the stable component during an observation, but changes every observation depending on the total intensity, $\langle I_\mathrm{in}\rangle $. 
$\lambda^{(n)}_\mathrm{opt}$ is the leakage coefficient.
The most important component is $n=4$, which appears as the polarization signal. The demodulated polarization signal becomes
\begin{equation}\label{eq:CRHWP5}
d_{\bar{d}}(t) = \varepsilon [ Q_\mathrm{in}(t) + {i} U_\mathrm{in}(t) ] + A_{0|\langle I_\mathrm{in}\rangle}^{(4)} + \lambda^{(4)}_\mathrm{opt} \delta I_\mathrm{in}(t) + \mathcal{N}_{\bar{d}}^{(\mathrm{Re})}(t) + {i} \mathcal{N}_{\bar{d}}^{(\mathrm{Im})}(t)\:.
\end{equation}
For an optical system with optical elements between the CRHWP and the sky,
$A_{0|\langle I_\mathrm{in}\rangle}^{(4)}$ and $\lambda^{(4)}_\mathrm{opt}$ mainly come from the instrumental polarization and the \ac{I2P} leakage, respectively. 
The non-idealities of the HWP, 
especially a mismatch of the \acf{ARC}
due to its birefringence,
mainly create $A_{0|\langle I_\mathrm{in}\rangle }^{(2)}$ and $\lambda^{(2)}_\mathrm{opt}$,
but it can also create small $A_{0|\langle I_\mathrm{in}\rangle}^{(4)}$ and $\lambda^{(4)}_\mathrm{opt}$ from the non-zero incident angle~\citep{ABS2016Systematics} or non-uniformity of the
\ac{ARC}.

In the case of the configuration shown in \autoref{fig:PB1_Xsec}, 
which is described in the next section,
the optical element between the CRHWP and the sky is the aluminum primary mirror~\citep{Tran2008}.
The instrumental polarization due to reflection by a metal surface is calculated as~\citep{MetalReflection}
\begin{equation}\label{eq:polarizedemission}
A_{0|\langle I_\mathrm{in}\rangle}^{(4)} = -\lambda^{(4)}_\mathrm{opt}[T_\mathrm{mirror}-\langle I_\mathrm{in}\rangle ]\sim60\,\mathrm{mK}\:,
\end{equation}
where $T_\mathrm{mirror}\sim300\,\mathrm{K}$ is the temperature of the primary mirror and $\langle I_\mathrm{in}\rangle \sim15\,\mathrm{K}$ is the total intensity from the sky.
The \ac{I2P} leakage coefficient of $n=4$ is expressed as~\citep{MetalReflection}
\begin{equation}\label{eq:mirrorleakage}
\lambda^{(4)}_\mathrm{opt} = -\varepsilon \sqrt{4\pi\epsilon_0\nu\rho}\,[\sec\chi-\cos\chi]\sim - 0.02\mathrm{\%}\:,
\end{equation}
where $\nu$ is the frequency of the incoming radiation, $\rho$ is the resistivity of the metal and $ \chi $ is the incident angle.\footnote{We ignore curvature of the mirror here.}
Here, the polarization modulation efficiency, $\varepsilon$ is applied since this signal is polarized. 
The expected values are estimated with $\varepsilon=1$, $\nu=148\,\mathrm{GHz}$~\citep{POLARBEAR2014ClBB}, $\rho=2.417\times10^{-8}\,\mathrm{\Omega}\cdot\mathrm{m}$~\citep{AlResistivity} for aluminum, and $\chi=32.5^\circ$ which measured from \autoref{fig:PB1_Xsec}.

Note that both $A_{0|\langle I_\mathrm{in}\rangle}^{(4)}$ and $\lambda^{(4)}_\mathrm{opt}$ are complex values, whose arguments represent the polarization angles.
In the off-axis optical systems as shown in \autoref{fig:PB1_Xsec}, these polarization angles are aligned to the optical plane and are almost uniform across the focal plane.

\subsection{Leakage from non-linearity of detector}
The non-idealities of the detector are another potential source of the \ac{I2P} leakage coupling with a non-zero amplitude of the $A_{0|\langle I_\mathrm{in}\rangle}^{(4)}$. 
In practice,
the detector\footnote{
In the case of \textsc{Polarbear}, we have
\ac{TES} 
bolometers, which are commonly used among recent and future CMB polarization experiments.} 
has a responsivity and a time-constant which change depending on the incoming total intensity and the focal plane temperature.
We calibrate them before and after an observation (see \autoref{sec:observation}),
but we cannot trace their variations during the observation.\footnote{Since the calibration tool may induce noise, it is usually turned off during the science observation.} 
In the small signal approximation, the responsivity and time-constant variations are proportional to the total intensity fluctuation. 
To first order, this induced non-linearity modifies the modulated signal $d_m(t)$ in \autoref{eq:d_m} as follows:
\begin{equation}\label{eq:d'}
d'_m(t) = [1 + g_1 d_m(t)]d_m(t-\tau_1 d_m(t)) \:,
\end{equation}
where $g_1 d_m(t)$ and $\tau_1 d_m(t)$ are variations of the responsivity and time constant.
Including the detector non-linearity, the observed intensity and polarization signals are modified from \autoref{eq:d0f} and \autoref{eq:CRHWP5} as follows,
\begin{align}
d'_{\bar{m}}(t) &= {\delta}I_\mathrm{in}(t) + \mathcal{N}_{\bar{m}}(t) \:,\label{eq:d'_m}\\
d'_{\bar{d}}(t) &= \varepsilon [ Q_\mathrm{in}(t) + {i} U_\mathrm{in}(t) ] + A_{0|\langle I_\mathrm{in}\rangle}^{(4)} + \lambda^{(4)} \delta I_\mathrm{in}(t) + \mathcal{N}_{\bar{d}}^{(\mathrm{Re})}(t) + {i} \mathcal{N}_{\bar{d}}^{(\mathrm{Im})}(t) \:,\label{eq:d'_d}
\end{align}
with
\begin{equation}\label{eq:totalleakage}
\lambda^{(4)} \equiv \lambda^{(4)}_\mathrm{opt} + 2 g_1 A_{0|\langle I_\mathrm{in}\rangle}^{(4)} + {i} \omega_\mathrm{mod} \tau_1 A_{0|\langle I_\mathrm{in}\rangle}^{(4)} \:.
\end{equation}
Here, parameters $g_1$, $\omega_\mathrm{mod}\tau_1$ and $\lambda^{(4)}_\mathrm{opt}$ are assumed to be small so that only first order terms are kept. 
\autoref{eq:totalleakage} shows that the total leakage, $\lambda^{(4)}$, could come not only from the optical leakage, $\lambda^{(4)}_\mathrm{opt}$, but also from 
detector non-linearity.

\section{Measurements of \ac{I2P} leakage}
\label{sec:measurements}

\subsection{Instrument}

\begin{figure}
\begin{center}
\begin{tabular}{lr}
\begin{minipage}{0.5\hsize}
\includegraphics[width=\textwidth]{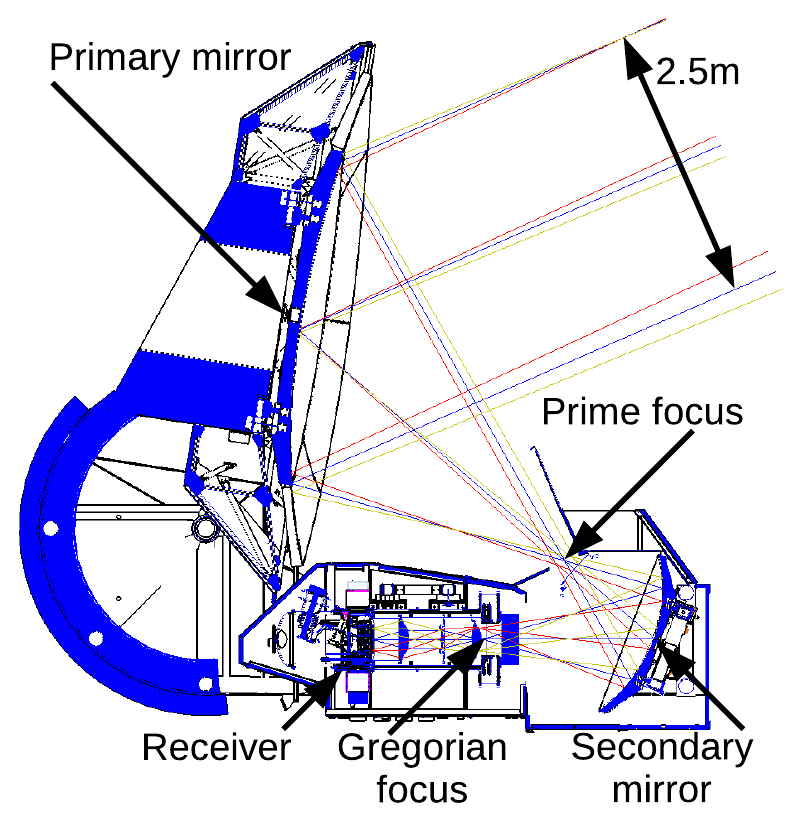}
\end{minipage}&
\begin{minipage}{0.45\hsize}
\includegraphics[width=\textwidth]{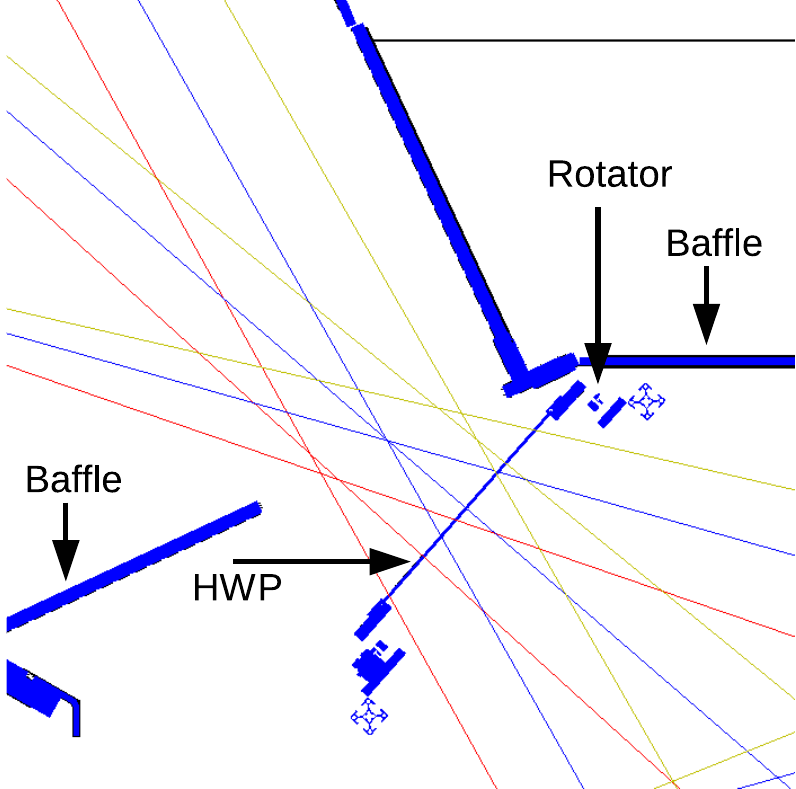}
\end{minipage}
\end{tabular}
\caption{Cross section of \pb (left) and 
detail of the area around the prime focus (right).}
\label{fig:PB1_Xsec}
\end{center}
\end{figure}

\textsc{Polarbear}~\citep{POLARBEAR2012SPIE} is a ground-based CMB polarization experiment in Chile.
The telescope has monolithic primary and secondary mirrors\footnote{The diameter of the major axis of the primary (secondary) is $2.96\,(1.47)\,\mathrm{m}$ and that of the minor axis is $2.93\,(1.38)\,\mathrm{m}$.}
in an off-axis Gregorian optics configuration employing a $2.5\,\mathrm{m}$ diameter aperture.
It has two foci, the prime focus and the Gregorian focus (\autoref{fig:PB1_Xsec}).
The Gregorian focus is re-imaged to the telecentric focal plane by re-imaging optics which consist of three cryogenic lenses in the \pb receiver.
Seven device wafers are placed at the focal plane and cooled to $250\,\mathrm{mK}$.
Each device wafer consists of 91 lenslet-coupled pixels and each pixel has an orthogonal polarization sensitive \ac{TES} bolometer pair.
The total of 1,274 \ac{TES} bolometers absorb the incoming optical power in a band centered on 148 GHz and convert the signal into current, which is amplified by SQUID amplifiers. The average instantaneous array sensitivity in the intensity signal during the first season of observations, $\mathrm{NET}_\mathrm{array}$, 
was $23\,\mu\mathrm{K}\sqrt{\mathrm{s}}$~\citep{POLARBEAR2014ClBB}.

A prototype CRHWP was installed at a position $11\,\mathrm{cm}$ behind the prime focus because of interference with the prime-focus baffle (right panel of \autoref{fig:PB1_Xsec}).
The HWP is made from a $28\,\mathrm{cm}$ diameter $3.1\,\mathrm{mm}$ thick single sapphire plate, which is coated for anti-reflection on both surfaces by a $0.23\,\mathrm{mm}$ thick layer of RT/duroid 6002 material.
The HWP is mounted on a rotator, which is a large ring gear supported by three small bearings. 
A stepper motor attached by a pinion gear spins the rotator at $\omega_\mathrm{HWP}/(2\pi)=2.0\,\mathrm{Hz}$ which produces a polarization signal modulated at $\omega_\mathrm{mod}/(2\pi)=8.0\,\mathrm{Hz}$.
These systems are operated at room temperature.
The relative angle of the HWP, $\theta_\mathrm{enc}(t) \equiv \theta_\mathrm{HWP}(t) - \theta_\mathrm{0}$, from an origin, 
$\theta_\mathrm{0}$,
is determined with 0.1${}^\circ$ accuracy from the encoder on the stepper motor and the 6:1 gear ratio between the motor and the HWP rotator.
The signal is sampled by a data acquisition system synchronized with the detector readout. 

Note that the rotator described above was not designed to be operated for year-long observations.
For our science observations using the CRHWP, 
which continued after the evaluation reported here,
we have developed and installed a new rotator which is more robust.
The results presented in this paper remain unchanged with the new rotator;
i.e.,
the size, fabrication process and the position of the CRHWP are the same, and the values of $A_{0|\langle I_\mathrm{in}\rangle}^{(4)}$ and $\lambda^{(4)}_\mathrm{opt}$ are consistent with those reported here. 
The details of the new rotator will be described in a forthcoming paper along with our scientific results.

\subsection{Observations}\label{sec:observation}
\pb started science observations in May 2012.
We observed three small patches targeting the lensing B-mode signal without the CRHWP.
The first observation using the CRHWP was done in February 2014. 
In this paper, we use eight observations taken through the night on February 14th, 2014, six of which were CMB observations, one of which was a Tau~A observation and one of which was a Jupiter observation. 
The data were taken at several different azimuths and elevations following the sky rotation.
Each observation included a one-hour set of scans and two types of calibrations before and after the scan set.

The CMB data were acquired with \acp{CES}, in which we scanned the sky back and forth in azimuth at a constant elevation to keep the atmospheric loading on the detector as stable as possible,
and to distinguish the sky signal from the ground pickup. 
The scan velocity was kept at 0.3${}^\circ/\mathrm{s}$ on the sky except during turnarounds.
The scan width was 15${}^\circ$ on the sky and the CRHWP was operating at $2.0\,\mathrm{Hz}$.

The Tau~A and Jupiter observations were used to calibrate the absolute angle and transmission of the CRHWP, respectively. 
Tau~A, also known as the Crab Nebula, is linearly polarized and it was also used to calibrate the relative polarization angle of the detectors in the analysis of previous science observations~\citep{POLARBEAR2014ClBB}.
Jupiter, which is one of the brightest sources in our band, was also regularly observed to determine the beam shape. 
Tau~A and Jupiter were observed with raster scans, in which the elevation is kept constant across one stroke and is stepped by 2${}'$ between strokes.
The scan velocity was kept at 0.2${}^\circ/\mathrm{s}$ on the sky and the CRHWP was rotated at $2.0\,\mathrm{Hz}$.

In the first calibration method,
we used a device called the stimulator to identify optically-responsive detectors and to calibrate the properties of the detectors.
The stimulator is placed at the back side of the secondary mirror and injects a chopped optical signal from a thermal source, which is kept at $700\,{}^\circ\mathrm{C}$,
through a small hole on the secondary mirror.
The chopper frequencies are changed in six steps from $4\,\mathrm{Hz}$ to $44\,\mathrm{Hz}$.
From the frequency dependence of the amplitudes seen in the detector, we calibrated the responsivity and the time constant of each detector.
The rotation of the HWP was stopped during the stimulator calibration.

For the other calibration method, called the elevation nod (el-nod),
we moved the telescope up and down by $2^\circ$ in elevation.
This injects an intensity signal due to the thickness variation of the atmosphere.
The CRHWP was rotated at $2.0\,\mathrm{Hz}$ during the el-nod calibration.

These calibrations were done sequentially;
each calibration set before a scan set started from the stimulator followed by the el-nod. The order was reversed in the calibration after the scan set.

\subsection{Calibration and demodulation}\label{subsec:CalibrationandDemodulation}
Here we describe the methods of calibration and demodulation to obtain the calibrated intensity and polarization signals, which are denoted as $d'_{\bar{m}}(t)$ and $d'_{\bar{d}}(t)$ in \autoref{sec:CRHWP}.

We use stimulator data to obtain the responsivity and the time-constant of each detector, which are then used to calibrate el-nod and CMB scan data~\citep{POLARBEAR2014ClBB}.
The responsivity can be derived from the apparent amplitude of the constant signal from the stimulator, which is calibrated by planet observations.
The transmission of the CRHWP, which is placed at the sky side of the stimulator, changes the calibration of the stimulator signal relative to sky signals.
We have measured the transmission as 90-95\% from the Jupiter observation for each device wafer, and it is applied to the stimulator signal calibration.\footnote{\label{fn:absolutecalibration}
We still could have systematic uncertainties in absolute values due to the sidelobe of the beam, the transmission of atmosphere, etc. Their impacts were estimated in \cite{POLARBEAR2014ClBB} to be $4.1\%$, which could increase due to the CRWHWP. While their careful control is crucial for B-mode measurements, they have limited impact on the fraction of \ac{I2P} leakage studied in this work and therefore are not included in the error budget.} 
El-nod data are calibrated using the value obtained from a stimulator run in the same calibration set.
CMB scan data are calibrated using the mean value between stimulator runs before and after the scan set.

We then extract intensity and polarization components from the calibrated data, $d'_m(t)$ using the demodulation method~\citep{MAXIPOL2007HWP,ABS2014Demod}
as shown in \autoref{eq:demodmethod}.
The intensity component was obtained by applying a low-pass filter up to $3.8\,\mathrm{Hz}$ by convolving 
\ac{FIR}
window filters.
To extract the polarization component, 
we first apply a band-pass filter around the modulation frequency with a $\pm 3.8\,\mathrm{Hz}$ band, then multiply the demodulation function by $2 e^{4{i}\theta_\mathrm{enc}(t)}$, and apply a low-pass filter with the identical cut-off frequency.
After the low-pass filtering, we downsample the timestreams by a factor of 24, down to $7.95\,\mathrm{Hz}$.
The $3.8\,\mathrm{Hz}$ band (corresponding to a maximum multipole $\ell$ of 4,560) is selected as slightly less than twice the frequency of the CRHWP rotation,
$2.0\,\mathrm{Hz}$, with $0.2\,\mathrm{Hz}$ margin to cut the sidebands of $n=2$ HWPSS.

For the polarization component, there are three effects which need to be corrected. 
One is the time-constant of the detector, $\tau$, which delays the phase of the modulated polarization signal by $\sim 9^{\circ}$ and also decreases the response to the signal by $\sim 1\%$.
This effect is corrected from the polarization component by multiplying by the inverse of the effect,
$(1-{i}\omega_\mathrm{mod}\tau)$, where $\tau$ is calibrated with the stimulator data.
Another effect is the polarization angle rotation due to the detector angle, $\theta_\mathrm{det}$, and the origin of the encoder, $\theta_\mathrm{0}$, which is corrected by multiplying a factor, $e^{2{i}\theta_\mathrm{det}+4{i}\theta_\mathrm{0}}$,
where $\theta_\mathrm{det}$ is calibrated in the science observation~\citep{POLARBEAR2014ClBB} and $\theta_\mathrm{0}$ is determined from the Tau~A observation to keep its angle consistent. 
After the correction, the real part and imaginary part of the polarization signal become Stokes $Q$ and $U$ with respect to the global instrumental coordinates defined by $(\mathrm{El},\mathrm{Az})$.
The other is the polarization efficiency, $\varepsilon$, which will be considered in a future paper.

\subsection{Methods 
to measure leakage}\label{sec:methods}
Now we consider how to measure the \ac{I2P} leakage.
As shown in \autoref{eq:totalleakage},
the leakage comes not only from the optical effect, but also from the non-linearity of the detector.
To distinguish these two, we perform three methods :
\begin{description}
\item[Method A :]Using the average of the instrumental polarization\footnote{This method can be applied to optical systems in which all the optical elements between the CRHWP and the sky are reflective.}\\
In this method, we use the amplitude of the instrumental polarization, which mainly comes from the polarized emission of the room temperature mirror.\footnote{We assume that the contribution from the non-ideality of the HWP on $n=4$ \acs{HWPSS}, $A_{0|\langle I_\mathrm{in}\rangle}^{(4)}$, is small compared to that from the instrumental polarization, here. See \autoref{sssec:OpticalLeakage} for a discussion about this.}
Using \autoref{eq:polarizedemission}, we can obtain the optical leakage 
\begin{equation}
\lambda^{(4)}_\mathrm{opt} = -\frac{ \left\llangle A_{0|\langle I_\mathrm{in}\rangle}^{(4)}\right\rrangle}{\left\llangle T_\mathrm{mirror}-\langle I_\mathrm{in}\rangle \right\rrangle}\:,
\end{equation}
where the double angle bracket represents average among observations.\footnote{One observation can be a CMB scan set or an el-nod calibration, but we use only el-nod here to combine with the total intensity measurement.}
Here, the instrumental polarization, $A_{0|\langle I_\mathrm{in}\rangle}^{(4)}$, is obtained from the average of the polarization signal. The temperature of the mirror, $T_\mathrm{mirror}$, is measured using a thermometer, and the total intensity from the sky, $\langle I_\mathrm{in}\rangle$, can be obtained from the el-nod observation or external \acl{PWV} information.

\item[Method B :]Using the variation of the instrumental polarization\footnote{This method is 
performed in~\cite{ABS2016Systematics}.}\\
In this method, we use the variation of the instrumental polarization $A_{0|\langle I_\mathrm{in}\rangle}^{(4)}$ correlated with the total intensity $\langle I_\mathrm{in}\rangle $, which will vary depending on the observing elevation and/or weather.\footnote{We assume that 
$T_\mathrm{mirror}$ is stable compared to the variation of $\langle I_\mathrm{in}\rangle $, here. This assumption is consistent with the measurement (see \autoref{sec:methodAexp}). A possible effect from the $T_\mathrm{mirror}$ variation is discussed in \autoref{sec:implementation}.} 
Using the derivative of \autoref{eq:polarizedemission}, we can obtain the optical leakage from the slope of the correlation as
\begin{equation}
\lambda^{(4)}_\mathrm{opt} = \frac{\Delta A_{0|\langle I_\mathrm{in} \rangle}^{(4)}}{\Delta \langle I_\mathrm{in}\rangle }\:,
\end{equation}
where $\Delta$ represents variation among observations e.g. $\Delta \langle I_\mathrm{in}\rangle \equiv \langle I_\mathrm{in}\rangle- \llangle \langle I_\mathrm{in} \rangle \rrangle$.

\item[Method C :]Using the leakage in the timestream\footnote{This method is performed in~\cite{MAXIPOL2007HWP}.}\\
In this method, we take the correlation between the intensity and polarization timestreams for each observation. 
Using the derivatives of \autoref{eq:d'_m} and \autoref{eq:d'_d}, the slope of the correlation results in the total leakage as
\begin{equation}
\lambda^{(4)} = \frac{\delta d'_{\bar{d}}(t)}{\delta d'_{\bar{m}}(t)}\:,
\end{equation}
where $\delta$ represents variation during each observation: e.g. $\delta d'_{\bar{d}}(t) \equiv d'_{\bar{d}}(t) - \langle d'_{\bar{d}} \rangle$.
The total leakage, $\lambda^{(4)}$, includes both the effects from optical and the detector non-linearity.

\end{description}
Note that we need many observations to obtain the leakage in the method A or B, while do one observation in C.
In other words, the time scale is longer than the calibration period for the method A or B, while not for C. 
That is the reason why we see the effect from the non-linearity of the detector only in the method C.

\subsection{Results of leakage measurements}

In this section, we describe the measurements of the \ac{I2P} leakage in the \pb experiment.
First, we perform the methods A and B using the el-nod data.
Then, we perform the method C with the el-nod data and the CMB scan data.

\subsubsection{Methods A and B : $\lambda_\mathrm{opt}^{(4)}$ measurements using 
instrumental polarization}\label{sec:methodAexp}
To perform methods A and B described in \autoref{sec:methods}, 
we need the $n=4$ HWPSS, $A_{0|\langle I_\mathrm{in}\rangle}^{(4)}$, 
the total intensity from the sky, $\langle I_\mathrm{in}\rangle $, and the temperature of the mirror, $T_\mathrm{mirror}$.

\begin{figure}
\begin{center}
\includegraphics[width=0.5\textwidth]{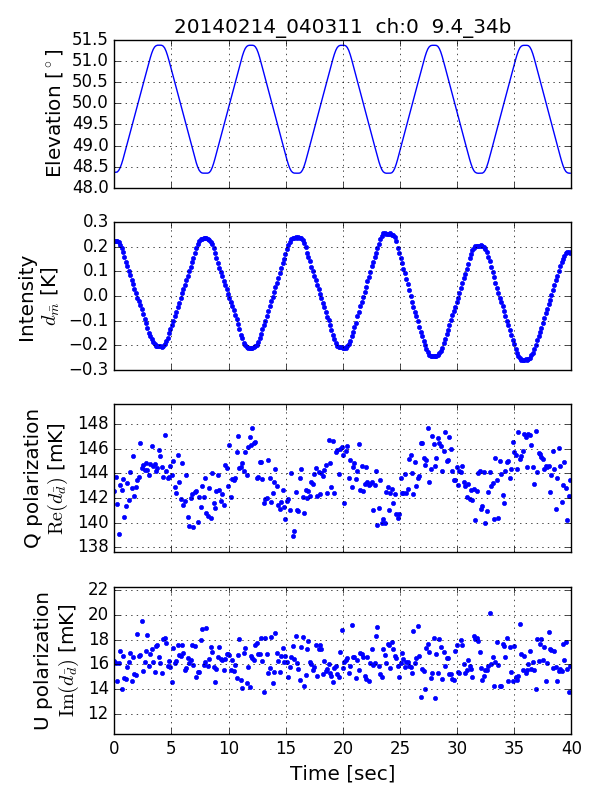}
\caption{Example \ac{TOD} 
from an el-nod observation. Top panel shows observing elevation. Second panel shows intensity signal. Third panel shows the real part of the polarization signal. Fourth panel shows the imaginary part of the polarization signal. The intensity signal depends on the thickness of the atmosphere along the line of sight combined with a small 1/f noise component.
The real part of the polarization signal shows variation from \ac{I2P} leakage above the white noise.}
\label{fig:example_elnod}
\end{center}
\end{figure}
The former two can be obtained from the el-nod observation.
\autoref{fig:example_elnod} shows example \acf{TOD} in an el-nod observation. 
The $n=4$ HWPSS, 
$A_{0|\langle I_\mathrm{in}\rangle}^{(4)}$, is obtained as the mean value of the polarization timestream, $\langle d'_{\bar{d}} \rangle$ (see \autoref{eq:d'_d}). 
The total intensity from the atmosphere, on the other hand, cannot be measured as the average of the intensity timestream, since the detector is only sensitive to the relative variation and there is larger uncertainty in the absolute value.
Here, we model the total intensity from the sky as follows:
\begin{equation}\label{eq:atmospheremodel}
I_\mathrm{in}(t) = I_0 + T_\mathrm{atm}\frac{\csc\mathrm{EL}(t)}{\langle{\csc\mathrm{EL}}\rangle}\:,
\end{equation}
where, $I_0$ is the contribution from the CMB, which is $1.6\,\mathrm{K}_\mathrm{RJ}$ in Rayleigh-Jeans temperature,\footnote{We used the bandpass measured in the laboratory using a Fourier transform spectrometer~\citep{POLARBEAR2012SPIEKam}.}
and the second term is the elevation-dependent intensity from the atmosphere, which is proportional to the cosecant of the elevation, $\mathrm{EL}(t)$.
The average and fluctuation of the total intensity become
\begin{equation}
\langle I_\mathrm{in}\rangle = I_0 + T_\mathrm{atm}\:,
\end{equation}
and
\begin{equation}\label{eq:elnodfluctuation}
\delta I_\mathrm{in}(t)\equiv I_\mathrm{in}(t)-\langle I_\mathrm{in}\rangle = T_\mathrm{atm}\left[\frac{\csc\mathrm{EL}(t)}{\langle{\csc\mathrm{EL}}\rangle}-1\right]\:,
\end{equation}
respectively.
The average of the total intensity from the atmosphere, $T_\mathrm{atm}$, can be obtained by fitting the correlation between the elevation, $\mathrm{EL}(t)$, and the intensity timestream, $d'_{\bar{m}}(t)=\delta I_\mathrm{in}(t)$, shown in \autoref{fig:example_elnod} with \autoref{eq:elnodfluctuation}.
Note that $A_{0|\langle I_\mathrm{in}\rangle}^{(4)}$ and $T_\mathrm{atm}$ are obtained for each detector for each el-nod observation.
The $T_\mathrm{mirror}$ is measured by a 1-wire digital thermometer on the primary mirror. Since the observations were done in the night, the $T_\mathrm{mirror}$ was stable, which was $270.6\,\mathrm{K}$ with $\pm 0.3\,\mathrm{K}$ variation at a maximum. 

 \begin{figure}\begin{center}
\begin{tabular}{cc}
\begin{minipage}{0.5\hsize}
\includegraphics[width=\textwidth]{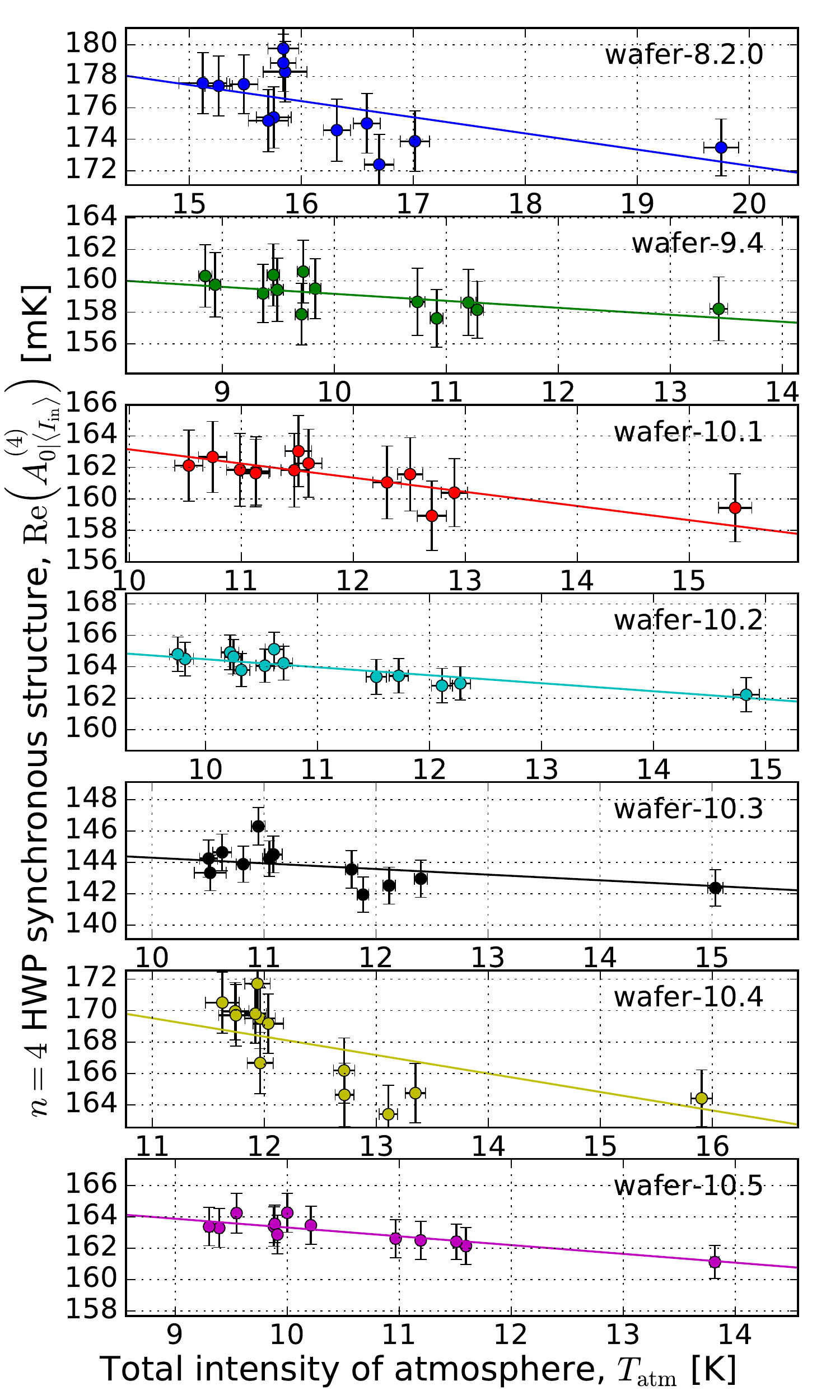}
\end{minipage}
\begin{minipage}{0.5\hsize}
\includegraphics[width=\textwidth]{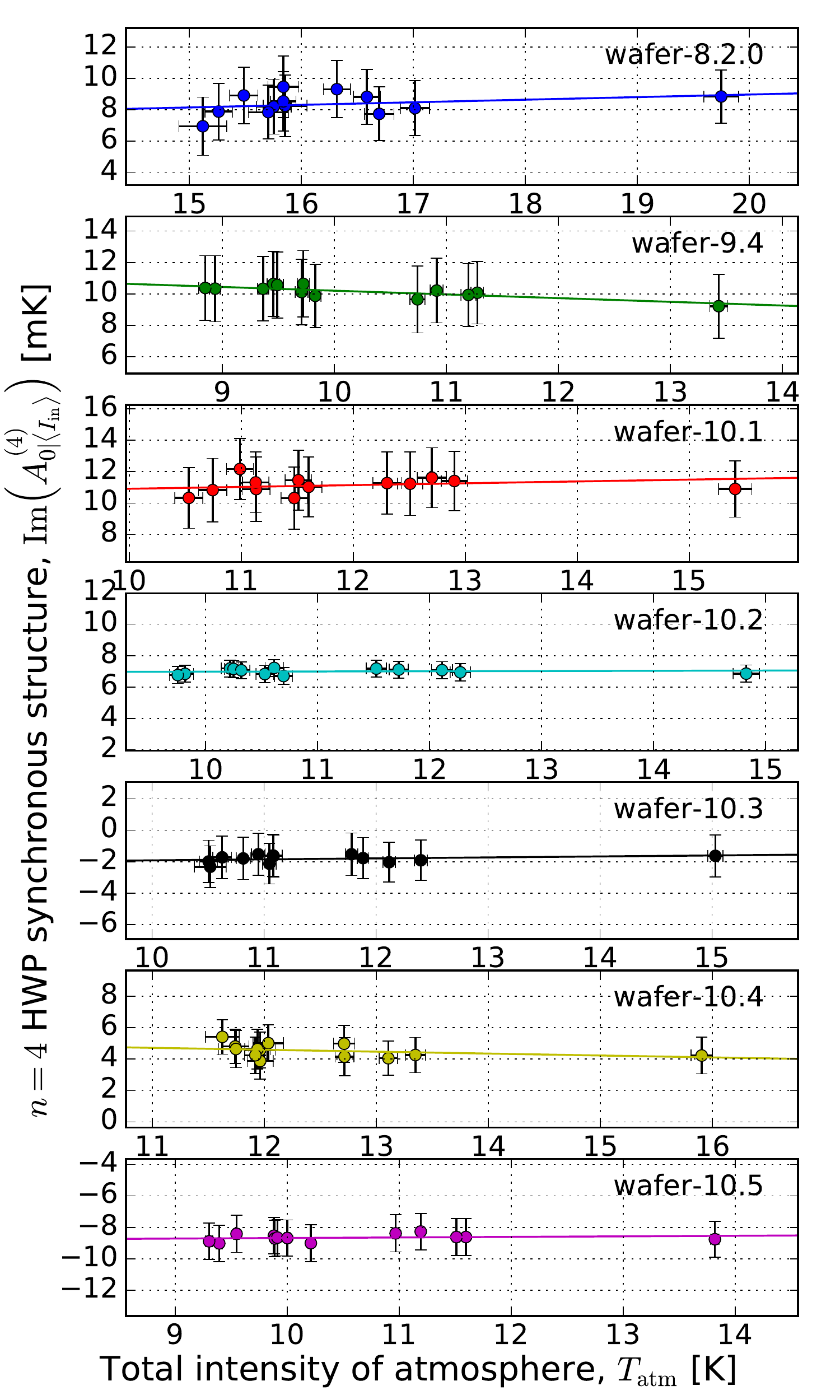}
\end{minipage}
\end{tabular}
\caption{\label{fig1}Correlation between the real (or imaginary) part of the HWPSS and the intensity from atmosphere is shown as left (or right) panel. 
Each point is obtained from each el-nod observation. It shows the average and one sigma error bar among all working detectors. 
The error bars show statistical uncertainties{\textsuperscript{\ref{fn:absolutecalibration}}}
dominated by the detector to detector systematic variation,
which is common among observations.
The difference in $T_\mathrm{atm}$ among the device wafers is due to the difference in detector frequency bandpasses between wafers. 
}
\end{center}
\end{figure}

The correlation between the real or imaginary part of the $n=4$ HWPSS and the total intensity from the atmosphere is shown in \autoref{fig1} for each device wafer.
Here, each point represents the average value for all detectors for each el-nod observation.
\autoref{fig1} also shows the best fit line of the correlation,
whose amplitude is listed in \autoref{tab1} with the optical leakage $\lambda^{(4)}_\mathrm{opt}$ which is derived using the method A,
and whose slope, which directly represents $\lambda^{(4)}_\mathrm{opt}$ in the method B, is listed in \autoref{tab2}. 
Slopes of the real part show small negative correlation for all device wafers from $-$0.04\% to $-$0.12\%.
Slopes of the imaginary part are smaller than slopes of the real part and have both positive and negative values.

Note that in the error bars in \autoref{fig1} or uncertainties in \autoref{tab1}, we considered five types of uncertainties\footnote{
The error bars and uncertainty values are the quadratic mean of the five uncertainties listed.
}
: uncertainty of the signal from each el-nod data, uncertainty in the responsivity from each stimulator calibration, uncertainty in the time-constant from stimulator calibration, statistical uncertainty in the absolute responsivity from previous science observation,{\textsuperscript{\ref{fn:absolutecalibration}}} and detector to detector variation.
For fitting the slope, however, we can remove the latter two uncertainties.
Since uncertainty in the absolute responsivity affects both $A_{0|\langle I_\mathrm{in}\rangle}^{(4)}$ and $T_\mathrm{atm}$ by the same ratio, the slope does not change.
Detector to detector variation can also be removed by subtracting the average for each detector, which also does not change the slope.
\begin{table}
\begin{center}
\begin{tabular}{|lrrrr|}
\hline
\parbox[b]{7.5em}{Device wafer\\(position on sky)} \rule[0pt]{0pt}{5ex}& $\mathrm{Re}\left(\left\llangle A_{0|\langle I_\mathrm{in}\rangle}^{(4)}\right\rrangle\right)$ & $\mathrm{Im}\left(\left\llangle A_{0|\langle 
I_\mathrm{in}\rangle}^{(4)}\right\rrangle\right)$ & $\mathrm{Re}\left(\lambda^{(4)}_\mathrm{opt}\right)$ (\%) & $\mathrm{Im}\left(\lambda^{(4)}_\mathrm{opt}\right)$ (\%)\\
\hline
8.2.0 (upper left)& $176.2\pm0.5\,\mathrm{mK}$ & $8.4\pm0.5\,\mathrm{mK}$ & $-0.0697\pm0.0002$ & $-0.0033\pm0.0002$ \\
9.4 (bottom)& $159.1\pm0.5\,\mathrm{mK}$ & $10.1\pm0.6\,\mathrm{mK}$ & $-0.0615\pm0.0002$ & $-0.0039\pm0.0002$ \\
10.1 (lower left)& $161.4\pm0.6\,\mathrm{mK}$ & $11.1\pm0.5\,\mathrm{mK}$ & $-0.0628\pm0.0002$ & $-0.0043\pm0.0002$ \\
10.2 (center)& $163.9\pm0.3\,\mathrm{mK}$ & $7.0\pm0.1\,\mathrm{mK}$ & $-0.0636\pm0.0001$ & $-0.0027\pm0.0001$ \\
10.3 (upper right)& $143.8\pm0.3\,\mathrm{mK}$ & $-1.8\pm0.4\,\mathrm{mK}$ & $-0.0558\pm0.0001$ & $0.0007\pm0.0001$ \\
10.4 (top)& $167.7\pm0.5\,\mathrm{mK}$ & $4.5\pm0.3\,\mathrm{mK}$ & $-0.0654\pm0.0002$ & $-0.0018\pm0.0001$ \\
10.5 (lower right)& $163.0\pm0.3\,\mathrm{mK}$ & $-8.7\pm0.3\,\mathrm{mK}$ & $-0.0631\pm0.0001$ & $0.0034\pm0.0001$\\
\hline
\end{tabular}

\caption{\label{tab1} The amplitude of the HWPSS 
obtained from the fit shown in \autoref{fig1} 
and the optical leakage from the method A for each device wafer. The uncertainties include only statistical contributions. Note that we have larger systematic uncertainties due to the absolute responsivity calibration.{\textsuperscript{\ref{fn:absolutecalibration}}}}

\end{center}
\end{table}
\begin{table}
\begin{center}
\begin{tabular}{|lrr|}
\hline
Device wafer & $\mathrm{Re}\left(\lambda^{(4)}_\mathrm{opt}\right)$ [\%] & $\mathrm{Im}\left(\lambda^{(4)}_\mathrm{opt}\right)$ [\%]\\
\hline
8.2.0 & $-0.103\pm0.008$ & $0.016\pm0.006$ \\
9.4 & $-0.044\pm0.006$ & $-0.024\pm0.003$ \\
10.1 & $-0.090\pm0.004$ & $0.012\pm0.007$ \\
10.2 & $-0.051\pm0.005$ & $0.001\pm0.001$ \\
10.3 & $-0.036\pm0.004$ & $0.006\pm0.002$ \\
10.4 & $-0.118\pm0.010$ & $-0.012\pm0.004$ \\
10.5 & $-0.056\pm0.003$ & $0.004\pm0.002$ \\
\hline
\end{tabular}

\caption{\label{tab2}The coefficient of the optical leakage 
obtained from the fit shown in \autoref{fig1} for each device wafer.}
\end{center}
\end{table}

\subsubsection{Method C : $\lambda^{(4)}$ measurement using 
el-nod data}\label{sec:methodBelnod}
Here, using method C described in \autoref{sec:methods}, we evaluate the total leakage, $\lambda^{(4)}$, which includes both the optical leakage and the leakage from the non-linearity of the detector.
We use el-nod data in this section.

Here, we obtain the coefficient of the leakage as follows to remove the bias from the white noise of the detector.
First we calculate the covariance matrix among the intensity signal and one of the polarization signals.
For the real part, for example, we obtain
\begin{align}
\mathrm{Cov}^{\mathrm{(Re)}} &\equiv \left (
\begin{array}{cc}
\langle d_{\bar{m}}^{\prime 2}\rangle & \langle d'_{\bar{m}}\mathrm{Re}(d'_{\bar{d}})\rangle /\sqrt{2} \\
\langle d'_{\bar{m}}\mathrm{Re}(d'_{\bar{d}})\rangle /\sqrt{2} & \langle \mathrm{Re}(d'_{\bar{d}})^{2}\rangle /2 
\end{array}
\right )\nonumber\\
&\approx \left(
\begin{array}{cc}
V_{I} + N & \mathrm{Re}(\lambda^{(4)}) V_{I} / \sqrt{2} \\
\mathrm{Re}(\lambda^{(4)}) V_{I} / \sqrt{2} & \mathrm{Re}(\lambda^{(4)})^2V_{I}/2 + N
\end{array}
\right)\:,
\end{align}
where $d'_{\bar{m}}(t)$ and $d'_{\bar{d}}(t)$ are the timestreams of an el-nod observation, 
$V_{I}\equiv\langle{\delta I_\mathrm{in}^2}\rangle$ is the variance of the intensity signal that mainly comes from the atmospheric fluctuation,
and $N$ is from the white noise of the detector (see \autoref{sec:modulation}).
Note that the polarization timestream is divided by a factor of $\sqrt{2}$ so that the white-noise terms in the diagonal components are same as $N$. 
Cross correlations of the white noise with other noise sources are assumed to be zero.
The eigenvalues and eigenvectors of this covariance matrix are calculated as follows:
\begin{align}
E_1^{\mathrm{(Re)}}&=[1+\mathrm{Re}(\lambda^{(4)})^2/2]V_{I} + N \:,\\
E_2^{\mathrm{(Re)}}&=N \:,\\
\overrightarrow{v_1}^{\mathrm{(Re)}}
&\propto (1,\mathrm{Re}(\lambda^{(4)}) / \sqrt{2})^\mathrm{T}\:,\\
\overrightarrow{v_2}^{\mathrm{(Re)}}
&\propto (-\mathrm{Re}(\lambda^{(4)}) / \sqrt{2},1)^\mathrm{T}\:,
\end{align}
where $E_1^{\mathrm{(Re)}}$ and $E_2^{\mathrm{(Re)}}$ are the eigenvalues and $\overrightarrow{v_1}^{\mathrm{(Re)}}$ and $\overrightarrow{v_2}^{\mathrm{(Re)}}$ are their eigenvectors.
We obtain the leakage coefficient from the ratio of the components of $\overrightarrow{v_1}^{\mathrm{(Re)}}$.
Its uncertainty is estimated using the eigenvalues and the number of samples, 
$n_\mathrm{sample}$, as $\sqrt{2} (E_1^{\mathrm{(Re)}}/E_2^{\mathrm{(Re)}})^{-1/2}n^{-1/2}_\mathrm{sample}$.
With a similar calculation for the imaginary component, we obtain $\mathrm{Im}(\lambda^{(4)})$, too.

Note that any additional components, such as a base temperature fluctuation and electrical noise, are not considered here, 
which might appear in the intensity-intensity component of the covariance matrix as $V_I + N + V_\mathrm{other}$. 
With this additional term, we would underestimate both the leakage and its uncertainty by a factor of $(1+V_\mathrm{other}/V_I)^{-1}$,
which may result in a residual 1/f noise (see \autoref{sec:implementation}).

The measured total leakage, 
which is averaged among observations and detectors for each device wafer, 
is shown in \autoref{fig:total_leakage_corr}.
We find that the absolute values for both the real part and imaginary part of the total leakage are much larger than those for the optical leakage obtained in the previous section; the real (imaginary) part of the total leakage is about 0.3\%~(0.1\%) for median and 0.8\%~(0.2\%) for maximum.

\subsubsection{Method C : $\lambda^{(4)}$ measurement using 
CMB scan data}\label{sec:methodBces}In the previous section, the intensity timestream, $d'_{\bar{m}}(t)$, is obviously dominated by the variation of the total intensity from atmosphere, $\delta I_\mathrm{in}(t)$.
In the case of hour long CMB scan data, however, $d'_{\bar{m}}(t)$ is dominated by the 1/f noise, which should come from the intensity fluctuation from the atmospheric turbulence, but may come from other sources.
Here, we perform method C again but with CMB scan data to check whether the leakage coefficient is consistent with that obtained from el-nod data.

Although the CMB scan is different from the el-nod (e.g.,
the data length is one hour for a CMB scan and 40 seconds for an el-nod),
we can use the exact same process explained in the previous section.

\begin{figure}\begin{center}

\includegraphics[width=\textwidth]{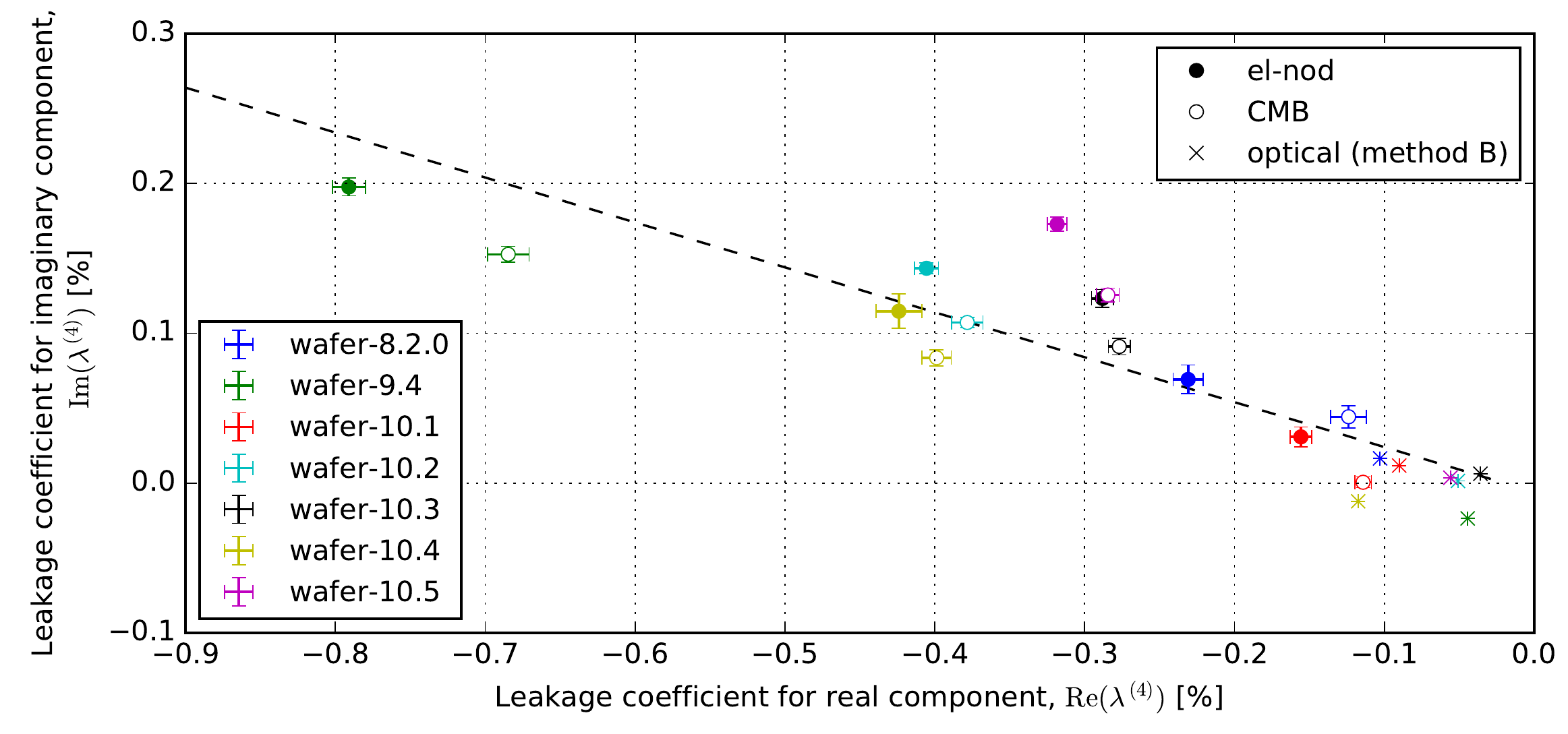}

\caption{\label{fig:total_leakage_corr}The total leakage coefficients from the 
el-nod and CMB scan data averaged among detectors on each device wafer. Each error bar shows the 1$\sigma$ uncertainty of the averaged value including the systematic uncertainty from detector-by-detector and observation-by-observation variation. The optical leakage coefficients are also plotted for comparison.}
\end{center}
\end{figure}

The total leakage measured from CMB scan data is also shown in \autoref{fig:total_leakage_corr}, which shows a similar trend with that obtained from el-nod data.

\subsection{1/f noise spectra}\label{sec:1/fnoiseeval}
Furthermore,
we check the noise spectra of CMB scan data to confirm that the correlation comes from a single mode, which should be the atmospheric fluctuation.
Since a single detector can measure both the intensity and polarization timestream at the same time as shown in \autoref{fig:example_elnod},
we can simply subtract the leakage as
\begin{equation}\label{eq:subtractedsignal}
d_{\bar{d}}^{(\mathrm{sub})}(t) \equiv d'_{\bar{d}}(t)-\lambda^{(4)}d'_{\bar{m}}(t) = \varepsilon [ Q_\mathrm{in}(t) + {i} U_\mathrm{in}(t) ] +A_{0|\langle I_\mathrm{in}\rangle}^{(4)} + \mathcal{N}_{\bar{d}}^{(\mathrm{Re})}(t) + {i} \mathcal{N}_{\bar{d}}^{(\mathrm{Im})}(t) - \lambda^{(4)}\mathcal{N}_{\bar{m}}(t)\:.
\end{equation}
\autoref{eq:subtractedsignal} shows no leakage term which contains $\delta I_\mathrm{in}(t)$, 
which means that we will not have any 1/f noise from atmosphere.
But if there is another source of 1/f noise, it will remain.
The last term means an increase of the white noise, but is negligible because of $|\lambda^{(4)}|\ll 1$. 

The noise spectra of the intensity timestream and the polarization timestreams,
before and after leakage subtraction, are calculated as follows.
To prevent mixing of spurious power in the discrete Fourier transformation from the gap between the edges of the data, we apply a second-order polynomial filter for the entire one-hour observation and apply a Hann window function.
Then \acp{PSD} are obtained by Fourier-transforming the intensity and polarization timestreams.
We fit the \acp{PSD}
with a 1/f noise spectrum formula:
\begin{equation}
P(f) = \left.P_\mathrm{1/f}\right|_{f_0}\left(\frac{f_0}{f}\right)^{\alpha}+P_\mathrm{white}
\end{equation}
where $f_0$ is an arbitrary pivot frequency, $\left.P_\mathrm{1/f}\right|_{f_0}$ is the PSD of the 1/f noise at the pivot frequency, $\alpha$ is the power law of the 1/f noise, and $P_\mathrm{white}$ is the PSD of the white noise.\footnote{The variance and PSD of the white noise are related by e.g. $P^{(\mathrm{int})}_\mathrm{white} = 2 N / f_\mathrm{sample}$ for the intensity signal, where $f_\mathrm{sample}$ is the sampling frequency of the data.} 
The frequency where the 1/f noise and the white noise become comparable, so called a knee frequency, is defined as
\begin{equation}
f_\mathrm{knee} \equiv \left(\frac{\left.P_\mathrm{1/f}\right|_{f_0}}{P_\mathrm{white}}\right)^{1/\alpha}f_0\:.
\end{equation}
Note that these procedures are applied to each detector.

We have to note, however, 
that atmospheric fluctuations are correlated among detectors in contrast to white noise.\footnote{The white noise is dominated by photon noise and phonon noise. Since detectors are not coherent, they are not correlated. } 
For structures with angular diameters larger than the field of view, the detector array behaves as a single detector with more than an order of magnitude better white-noise sensitivity than that of a single detector due to averaging.
Since the correlated 1/f noise is not suppressed by averaging, it may become significant in the PSD of the averaged data.
Besides, the 1/f noise spectra of the averaged timestream are more meaningful for observations at large angular scales.

The averaged polarization signal is estimated as
\begin{equation}
d'_{\bar{d},\mathrm{array}}(t)\equiv\left.\left(\sum_i \frac{d'_{\bar{d},i}(t)}{P_{\mathrm{white},i}^{(\mathrm{pol})}}\right)\right/\left(\sum_i \frac{1}{P_{\mathrm{white},i}^{(\mathrm{pol})}}\right)\:,
\end{equation}
where $i$ denotes the index of each detector, $d'_{\bar{d},i}(t)$ is the polarization timestream data of the detector, and $P_{\mathrm{white},i}^{(\mathrm{pol})}\equiv P_{\mathrm{white},i}^{(\mathrm{Re})}+P_{\mathrm{white},i}^{(\mathrm{Im})}$ is the sum of the white-noise PSD for the real and imaginary parts of the polarization signal, $d'_{\bar{d},i}(t)$. 
The averaged intensity signal, $d'_{\bar{m},\mathrm{array}}(t)$ is also obtained as the weighted average of the intensity signals among detectors, where $1/P_{\mathrm{white},i}^{(\mathrm{pol})}$ is used for weighting.\footnote{If PSD is dominated by 1/f noise,
as is the case of intensity, we cannot obtain the white-noise PSD.} 
The PSDs are evaluated in the same manner for each detector.

\begin{figure}
\begin{center}
\begin{tabular}{c}
\begin{minipage}{0.5\hsize}
\includegraphics[width=\textwidth]{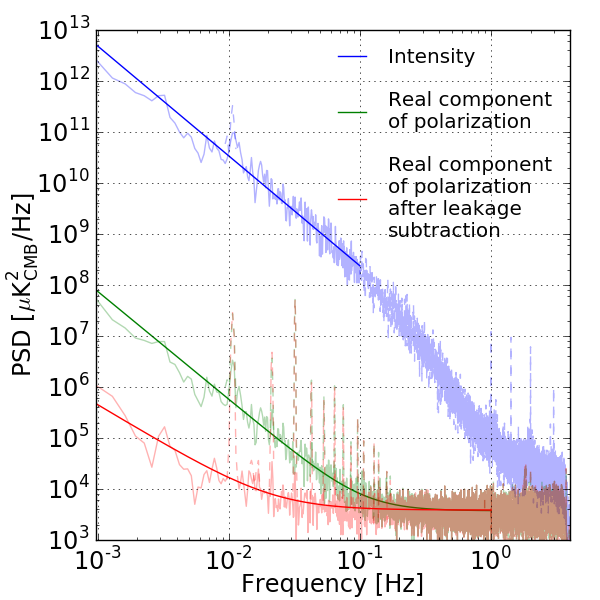}
\end{minipage}
\begin{minipage}{0.5\hsize}
\includegraphics[width=\textwidth]{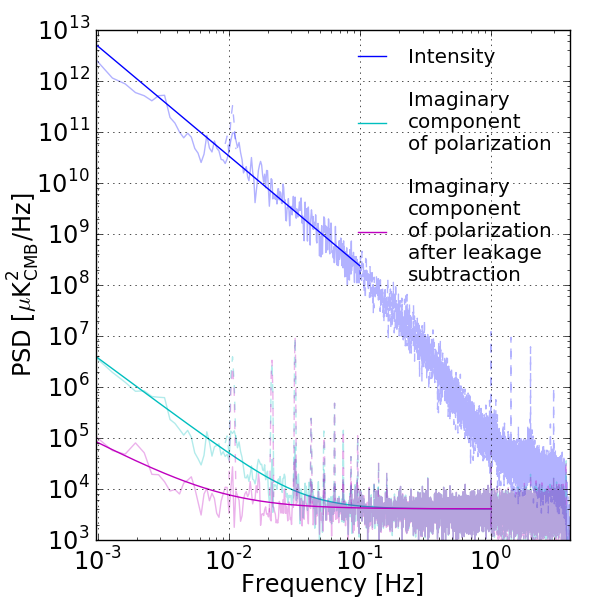}
\end{minipage}
\end{tabular}
\caption{\label{fig1/f}\acp{PSD} of coadded timestreams 
for all focal plane detectors for the real part (left panel) and for the imaginary part (right panel). The blue line shows the intensity fluctuation and the green (cyan) and red (magenta) lines show the real (imaginary) part of the polarization signal before and after leakage subtraction, respectively. 
The spikes at the harmonics of $0.01\,\mathrm{Hz}$ are the scan synchronous signals.}
\end{center}
\end{figure}

\autoref{fig1/f} shows \ac{PSD} of the averaged data for the intensity signal and polarization signal before and after 
leakage subtraction.
The fit curves with 1/f noise spectrum are also plotted.
We find that the 1/f noise in the polarization signal before leakage subtraction has decreased more than an order of magnitude in power after leakage subtraction.
The knee frequency has also improved by about an order of magnitude. 
This result suggests that the 1/f noise of intensity and polarization timesteams are dominated by the atmospheric intensity fluctuation and its leakage.

Note that peaks in the polarized signal at the harmonics of the scan frequency, $0.01\,\mathrm{Hz}$, are the scan synchronous signals, which come from the ground as observed by the far side-lobe of the telescope and could be somewhat polarized.
We subtract such scan synchronous signals by applying a ground-template filtering in map-making process~\citep{POLARBEAR2014ClBB}.

\section{Discussion and implementation}
\label{sec:discussion}
In this section, we discuss \ac{I2P} leakage measured in the previous section, comparing it with our expectations.
We also discuss the impact of this study on the measurements of cosmological parameters, in particular the tensor-to-scalar ratio,
from the aspect of uncertainties due to 1/f noise and the systematics.
Finally, we discuss the implications of this study for future CMB polarization experiments.

\subsection{\ac{I2P} leakage}
As described in \autoref{sec:CRHWP}, there are two origins of the \ac{I2P} leakage; one is the optical leakage and the other is the leakage due to detector non-linearities.
In the previous section, we have shown four types of measurements to shed light on the origin of the \ac{I2P} leakage.

\subsubsection{Optical leakage}\label{sssec:OpticalLeakage}
The optical leakage, $\lambda^{(4)}_\mathrm{opt}$,
measured using method B in \autoref{tab2} is found to be between $-$0.04\% and $-$0.12\%. 
The trend that the imaginary part of the optical leakage is closer to zero than the real part, agrees with the expectation.
With method A in \autoref{tab1}, we obtain $-0.06\%$ from all the device wafers, which is in agreement with results from method B.
The following properties seen in method A are also in agreement with the expectation:
the $n=4$ HWPSS, $A_{0|\langle{I_\mathrm{in}}\rangle}^{(4)}$, is polarized, and its polarization angle is aligned with the vertical direction of the telescope.
Our na\"{\i}ve expectation of the optical leakage is $\lambda^{(4)}_\mathrm{opt}\sim -0.02$\% from \autoref{eq:mirrorleakage},
which is close to our measurements but is smaller.
This can be attributed to the uncertainty in the property of the mirror material or the non-idealities of the HWP. 
Among device wafers, we could have $\pm8\%$ relative variation because of the difference in the incident angle $\chi$ by $\pm1^{\circ}$, and $\pm2\%$ relative variation from bandpass center frequency difference by $\pm5\,\mathrm{GHz}$, according to \autoref{eq:mirrorleakage}.
Besides, peripheral wafers might also have contributions from the diffraction at edge of the rotator and the non-uniformity of the ARC.
However, we cannot find any clear feature of these effects in \autoref{tab1} due to systematic uncertainty in calibration.

In summary, the measured amplitude of the optical leakage is at a level of about 0.1\% or less, and overall properties agree with the expectations.

\subsubsection{\ac{I2P} leakage from detector non-linearity}\label{sssec:DetectorNonideality}
The amplitude of the total leakage, $\lambda^{(4)}$, measured by method C in \autoref{sec:methodBelnod} and \autoref{sec:methodBces}, was found to be much larger than the optical leakage, $\lambda^{(4)}_\mathrm{opt}$, measured by method A or B.
Here we assume that the difference comes from the leakage due to the detector non-linearity, 
as described in \autoref{eq:totalleakage}, and assess the effect more quantitatively.

The non-linearity of the detector can be estimated from the physical model of the detector.
By expanding equations for the constant-voltage-biased \ac{TES} bolometer~\cite[e.g.][]{TES2005} up to second order perturbations, we obtain the non-linearity of the responsivity and time-constant as follows:
\begin{equation} \label{eq:modelg1}
g_1 \approx -\frac{\eta}{2P_\mathrm{elec}}\frac{\mathcal{L}}{\mathcal{L}+1}\frac{\mathcal{L}+1+\omega_\mathrm{mod}^2\tau_0^2}{(\mathcal{L}+1)^2+\omega_\mathrm{mod}^2\tau_0^2}C\:,
\end{equation}
and
\begin{equation} \label{eq:modeltau1}
\tau_1 \approx \tau_0\frac{\eta}{P_\mathrm{elec}}\frac{\mathcal{L}^2}{\mathcal{L}+1}\frac{1}{(\mathcal{L}+1)^2+\omega_\mathrm{mod}^2\tau_0^2}C\:,
\end{equation}
where $\mathcal{L}$ is the effective loop gain, $\eta$ is a conversion factor from kelvin to pico-watt and $P_\mathrm{elec}$ is the electrical power injected into the TES circuit, and $\tau_0=(\mathcal{L}+1)\tau$ is the thermal time constant of TES bolometer, and $C$ is a number related to the second order derivatives of the TES resistance model, $R(T,I)$. 
The design value of $\eta$ is about $0.18\,\mathrm{pW/K}$~\citep{ZiggyPhD}
but here we use $0.17\,\mathrm{pW/K}$ due to the transmission of the CRHWP. 
$P_\mathrm{elec}$ can be obtained from data as about $10\,\mathrm{pW}$.
The thermal time-constant $\tau_0$ is measured using several calibration data as about $10\,\mathrm{ms}$.
The loop gain is estimated from the time-constant and it is about two.
$C$ is estimated from two popular models, two-fluid model~\citep{Irwin1998} or resistively shunted junction (RSJ) model~\citep{Kozorezov2011}, as 0.83 or 2.1 at fractional resistance of 60\%.
Then, the non-linearity is estimated as $-0.17\%\mathrm{/K}$ or $-0.42\%\mathrm{/K}$.
Using the measured amplitude of the $n=4$ HWPSS of $0.16\,\mathrm{K}$,
the leakage from non-linearity is estimated as $2g_1 A_{0|\langle{I_\mathrm{in}}\rangle}^{(4)}=-$0.05\% or $-$0.13\%.
The time-constant drift, $\tau_1$, is estimated as $0.02\,\mathrm{ms/K}$ or $0.05\,\mathrm{ms/K}$, which corresponds to the imaginary part of the leakage as $\omega_\mathrm{mod}\tau_1 A_{0|\langle{I_\mathrm{in}}\rangle}^{(4)}=0.02\%$ or 0.04\%.

The trend of the device wafer difference in \autoref{fig:total_leakage_corr} 
agrees with the model.
Detectors on device wafer 9.4 are operated at lower electrical power, about $5\,\mathrm{pW}$, and show larger leakage.
On the other hand, those on the device wafer 10.1 are operated at larger electrical power, about $30\,\mathrm{pW}$, and show smaller leakage.
From \autoref{eq:modelg1}, \autoref{eq:modeltau1} and \autoref{eq:totalleakage}, 
the ratio of the leakage into real and imaginary parts can be estimated as
\begin{equation}\label{eq:leakageratio}
\frac{\omega_\mathrm{mod} \tau_1 A_{0|\langle{I_\mathrm{in}}\rangle}^{(4)}}{2 g_1 A_{0|\langle{I_\mathrm{in}}\rangle}^{(4)}} = -\frac{\mathcal{L}\omega_\mathrm{mod}\tau_0}{\mathcal{L}+1+\omega_\mathrm{mod}^2\tau_0^2}\sim -0.3\:,
\end{equation}
which is in good agreement with the measurement shown in \autoref{fig:total_leakage_corr}.

All the results mentioned above follow the same trend as the model.
Therefore, we suspect that the transition shape dependent factor, $C$, 
is responsible for the difference in the absolute values between the measurements and expectation.
The physical model of the TES transition is not yet fully understood, which is a source of uncertainties.
Besides, our TES is AC biased~\citep{Dobbs2012RSI}, which might have different properties from DC biased TES as pointed out in {{{van der Kuur} et~al.}}~\citep{Kuur2011}. 
If $C$ is about 4, 
expectations of both real and imaginary parts of the total leakage are in good agreement with the measured values.
The direct measurement of the TES parameters is needed for further investigation.

We have also investigated other origins of non-idealities, such as a load resistance, a stray inductance, and the SQUID amplifier, and found that the load resistance has the largest effect among these.
The load resistance is the sum of a $0.03\,\mathrm{\Omega}$ shunt resistance and a parasitic resistance in the TES circuit,
and is estimated to be about $0.1\,\mathrm{\Omega}$ or less compared to the operating TES resistance of $1.0\,\mathrm{\Omega}$.
Including the load resistance into the model changes the responsivity and increases the leakage into the real part from $-$0.05\% to $-$0.1\%, but it does not change the leakage into the imaginary part, which is determined by the time-constant. Then, the imaginary-to-real ratio of the estimated leakage decreases from \autoref{eq:leakageratio}, which does not agree with the data in
\autoref{fig:total_leakage_corr}. 

\subsubsection{Comparison with other experiments}
Here we compare our measurements of the instrumental polarization and the \ac{I2P} leakage with the MAXIPol and ABS experiments, 
and summarize in \autoref{tab:comparison}. 

\begin{table}[t]
\begin{center}
\begin{minipage}{\textwidth}
\renewcommand\footnoterule{}
\begin{tabular}{|l|l|r|r|r|r|}
\hline
Experiment & \parbox[b]{9em}{Position of CRHWP} & \parbox[b]{3.5em}{\centering$|A_{0|\langle I_\mathrm{in}\rangle}^{(4)}|$\\ (mK)} & \parbox[b]{4.5em}{\centering $|\lambda_\mathrm{opt}^{(4)}|$\\method A} & \parbox[b]{4.5em}{\centering $|\lambda_\mathrm{opt}^{(4)}|$\\method B} & \parbox[b]{4.5em}{\centering $|\lambda^{(4)}|$\\method C}\\\hline MAXIPol~\citep{MAXIPOL2007HWP} & \parbox[t]{9em}{After one warm mirror, one window, and two cold mirrors\vspace{1ex}} & $33 \mbox{--} 600$ & \rule[0.7ex]{4.5em}{0.1pt} & \rule[0.7ex]{4.5em}{0.1pt} & $\lesssim1\%\mbox{--}5\%$\\\hline ABS~\citep{ABS2016Systematics} & First optical element & $40$ & \rule[0.7ex]{4.5em}{0.1pt} & $\sim0.013\%$ & $<0.07\%$,\footnote{This is a conservative 2-$\sigma$ upper limit for the average across the focal plane.}\\\hline
\parbox[t]{5em}{\pb (this work)} & After the primary & $\sim160$ & $\sim0.06\%$ & $\lesssim0.12\%$ & $\lesssim0.9\%$\\\hline
\end{tabular}
\end{minipage}
\caption{\label{tab:comparison}Comparison of instrumental polarization and \ac{I2P} leakage}
\end{center}
\end{table}

MAXIPol~\citep{MAXIPOL2007HWP} has an optical system, which has a warm primary mirror, a polypropylene vacuum window, two cold mirrors and an aperture stop on the sky side of the CRHWP.
They found that $|A_{0|\langle{I_\mathrm{in}}\rangle}^{(4)}|$ ranges from $33\,\mathrm{mK}$ to $600\,\mathrm{mK}$ and $|\lambda^{(4)}|$ from $\lesssim1\%$ to $5\%$ for each detector, which was measured by method C using Jupiter observations.

ABS~\citep{ABS2016Systematics} adopts an optical system in which the CRHWP is the first optical element.
There is no instrumental polarization and therefore, $A_{0|\langle{I_\mathrm{in}}\rangle}^{(4)}$ and $\lambda^{(4)}_\mathrm{opt}$ have only small contributions from non-uniformity of the HWP.
They found $|A_{0|\langle{I_\mathrm{in}}\rangle}^{(4)}|$ $\sim$ $40\,\mathrm{mK}$ and $|\lambda^{(4)}_\mathrm{opt}| \sim 0.013\%$ for each detector, which was measured by method B. 
They also put an upper limit on the leakage from Jupiter observations as $|\lambda^{(4)}|<0.07\%$, which corresponds to method C.

In spite of differences in the optical system, such as the number of mirrors and the position of the CRHWP, our measurements of $|\lambda^{(4)}_\mathrm{opt}| \sim 0.06\%$ from method A or $|\lambda^{(4)}_\mathrm{opt}| < 0.12\%$ from method B and $|\lambda^{(4)}|<0.9\%$ from method C are similar with those measurements within an order of magnitude. 

\subsection{Impact on 
B-mode power spectrum measurements}
In this section, 
we discuss the impact of this study on the measurements of cosmological parameters, in particular the tensor-to-scalar ratio, $r$. 
The total leakage described in the previous section results in 1/f noise and systematic uncertainties for CMB B-mode angular power spectrum measurement. Both affect $\sigma(r)$ which is the total uncertainty on $r$.
Using the leakage subtraction described in \autoref{eq:subtractedsignal}, however, we can minimize the uncertainty.
\subsubsection{1/f noise}
As mentioned in e.g. \cite{ABS2014Demod} and shown in \autoref{eq:d'_d}, the \ac{I2P} leakage contaminates the polarization signal with the 1/f noise in the intensity.
As described in the previous section,
the leakage coefficient could be not only the optical leakage, $\lambda^{(4)}_\mathrm{opt}$, but the total leakage, $\lambda^{(4)}$, 
which includes the effects from detector non-linearities.
Without the leakage subtraction, 
in the real or imaginary part,
a 0.4\% or 0.1\% leakage causes the most significant contribution to the 1/f noise as shown in \autoref{fig1/f}.
Note that the 1/f noise due to \ac{I2P} leakage is correlated among detectors and does not decrease by averaging, as opposed to the white-noise, as explained in \autoref{sec:1/fnoiseeval}.

Simple template subtraction, using the intensity signal, 
seems to be able to reduce the leakage by an order of magnitude in the PSD.
By comparing the PSDs for the intensity signal and polarization signal after the leakage subtraction, we can set an upper limit on the \ac{I2P} leakage of $\lesssim 0.1\%$.
The performance is as good as the results from ABS~\citep{ABS2014Demod}.

We can na\"{\i}vely evaluate the 1/f noise in terms of multipole by scaling the frequency using the scan velocity of $0.3^\circ/\mathrm{s}$. 
For the real part (or imaginary part),
the knee frequency of the PSD after leakage subtraction is $22\,(9)\,\mathrm{mHz}$, which corresponds to a multipole $\ell\simeq 26 \,(11)$.
However, the knee frequency depends on the white noise which could change every observation due to the observation condition.
Assuming the typical array sensitivity, $\mathrm{NET}_\mathrm{array}$, of $23\,\mu\mathrm{K}\sqrt{\mathrm{s}}$ at \textsc{Polarbear}~\citep{POLARBEAR2014ClBB}, the knee frequency becomes $32\,(15)\,\mathrm{mHz}$, which corresponds to multipole $\ell \simeq 39\,(18)$.
This na\"{\i}ve estimation indicates promising 1/f noise performance for the measurement of $r$, whose signal becomes significant at $\ell<100$.
Note that to be precise the angular noise power spectrum has to be evaluated for two-dimensional maps, thus depends on scan strategies and map-making methods~\citep[see e.g.][]{Brown2009}. 
Such realistic evaluation of the 1/f noise performance lies outside the scope of this study, 
and will be reported in a future paper.

\subsubsection{Systematic uncertainties from the \ac{I2P} leakage}\label{sec:systematics}The \ac{I2P} leakage also contaminates the polarization signal with CMB temperature fluctuation, which could contribute to the systematic uncertainty for the CMB B-mode power spectrum measurement.
As mentioned in \cite{ABS2016Systematics}, this effect is equivalent to the differential gain systematics~\cite[e.g.][]{HHZ2003,Shimon2008}, which directly leads to leakage of $C_\ell^{TT}$ into $C_\ell^{EE}$ and $C_\ell^{BB}$ independent of the beam resolution.\footnote{Other beam systematics are kinds of spatial derivatives, which are suppressed by a factor of $(\ell/\ell_\mathrm{beam})^2$ or $(\ell/\ell_\mathrm{beam})^4$, where $\ell_\mathrm{beam}$ is inverse of the beam resolution.} 
Therefore, \ac{I2P} leakage could be problematic especially for precision measurements at large angular scales.\footnote{This systematic could be mitigated by the scan strategy and the {\it deprojection} method~\citep{BICEP2Systematics2015} to some extent.
}

Compared to the primordial B-mode polarization, the bias in the B-mode power spectrum from the $\sim 0.1\%$ leakage is a level of $r \sim 0.03$ at $\ell \sim 80$.
In practice,
the error can be reduced several orders of magnitude by e.g. sky rotation and cancellation with multiple observations.
Therefore, it would be possible to reduce this systematic uncertainty at a level of $r < 0.01$, which is sufficient for current experiments. 
More realistic estimation will be reported in a future paper using the science observation data. 

\subsection{Implications for future projects}\label{sec:implementation}

Future ground-based experiments, such as Simons Array~\citep{SimonsArray2016SPIE}, Advanced ACTPol~\citep{AdvACTPol2015LTD}, and CMB-S4~\citep{CMBS4} may benefit from using a CRHWP to access low multipoles with a large-aperture telescope.
For Simons Array, the position of the HWP will be at the Gregorian focus, since the cross-polarization performance is better than at the prime focus and a larger HWP has become available~\citep{Hill2016SPIE}. 
Additionally, a mechanism to rotate the HWP at cryogenic temperature is under development~\citep{EBEX2011rotator}.
However, at the Gregorian focus, there will be two mirrors between the HWP and the sky. 
Since each of them will have \ac{I2P} leakage and polarized emission, 
both the optical leakage and the $n=4$ HWPSS could be roughly twice that reported in this study.

On the other hand, the array sensitivity will be an order of magnitude better than the current \pb receiver.
We will need to determine the leakage coefficient at a precision of the order of 0.01\% in order to achieve the same 1/f knee frequency of this study and also to satisfy requirements of systematic uncertainties for measurements of $r\sim0.001$. 

It is also important to consider possible origins of the 1/f noise that remains after subtraction. 
It is likely that the residual 1/f noise is primarily due to fluctuations of the mirror temperature and/or focal plane temperature. 
These fluctuations may contaminate the intensity and polarization signals as
\begin{align}
d'_{\bar{m}}(t) =& \delta I_\mathrm{in}(t) + \delta I_\mathrm{base}(t) + \mathcal{N}_{\bar{m}}(t)\:, \\
d'_{\bar{d}}(t) =& \varepsilon [ Q_\mathrm{in}(t) + {i} U_\mathrm{in}(t) ] + A_{0|\langle{I_\mathrm{in}}\rangle}^{(4)} + \lambda^{(4)} \delta I_\mathrm{in}(t) + \delta P_\mathrm{mirror}(t) + \mathcal{N}_{\bar{d}}^{(\mathrm{Re})}(t) + {i} \mathcal{N}_{\bar{d}}^{(\mathrm{Im})}(t)\:,
\end{align}
where $\delta I_\mathrm{base}(\delta T_\mathrm{base}(t))$ and $\delta P_\mathrm{mirror}(\delta T_\mathrm{mirror}(t))$ are the equivalent intensity and polarization signal due to the focal plane temperature fluctuation, $\delta T_\mathrm{base}(t)$, and the mirror temperature fluctuation, $\delta T_\mathrm{mirror}(t)$, respectively.
Then, the leakage subtraction as \autoref{eq:subtractedsignal}
results in
\begin{align}\label{eq:residualnoise}
d_{\bar{d}}^{(\mathrm{sub})}(t) =& \varepsilon [ Q_\mathrm{in}(t) + {i} U_\mathrm{in}(t) ] + A_{0|\langle{I_\mathrm{in}}\rangle}^{(4)} + \mathcal{N}_{\bar{d}}^{(\mathrm{Re})}(t) + {i} \mathcal{N}_{\bar{d}}^{(\mathrm{Im})}(t) - \lambda^{(4)}\mathcal{N}_{\bar{m}}(t) \nonumber\\
& - \lambda^{(4)}\delta I_\mathrm{base}(t) + \delta P_\mathrm{mirror}(t)\:.
\end{align}
The last two terms of \autoref{eq:residualnoise} may cause residual 1/f noise.
Another possibility is the responsivity variation from sources other than detector non-linearity, such as readout electronics. It can be included in the model by changing the coefficient of \autoref{eq:d'} into $[1+g_1 d_m(t) + \delta g(t)]$, where $\delta g(t)$ is the additional responsivity variation. The new coefficient results in an additional noise term, $A_{0|\langle{I_\mathrm{in}}\rangle}^{(4)}\delta g(t)$ in $d'_{\bar{d}}(t)$. 
In the future experiment, 
better temperature control for mirrors, the focal plane, and the readout electronics is recommended.

Additionally, 
there is a strong argument to minimize the leakage coefficients by minimizing the detector non-linearity, as the systematic uncertainty from imperfect knowledge of the leakage coefficients degrades the B-mode power spectrum measurement.
As the detector non-linearity arises from resistance variation in the TES, 
we can reduce the detector non-linearity by operating under a higher loop gain.
Furthermore,
there are ideas to eliminate the detector non-linearity:
operating the SQUID amplifier with a digital active nulling~\cite{deHaan2012SPIE} and operating the TES in a resistance-locked loop~\citep{Kuur2013}.

\section{Summary}
\label{sec:summary}
The \ac{CRHWP} is one of the most promising tools to reduce the impact of low-frequency noise in polarization measurements with large aperture telescopes.
The optimal position for a CRHWP is skywards of the telescope, but size limitations on the HWP diameter rule this out for large-aperture telescopes. 
In \textsc{Polarbear},
we place the CRHWP in the middle of the optics chain. 
The optics located on the sky side of the CRHWP cause 
\ac{I2P} leakage and 
instrumental polarization.
Additionally, the instrumental polarization couples with the non-linearities of the detector, which results in another source of the \ac{I2P} leakage.
The total \ac{I2P} leakage including both the effects could become problematic as a source of low-frequency noise and systematics.

We examined the \ac{I2P} leakage of a CRHWP at the prime focus of the \pb experiment using three methods. 
We measured the amplitude of the optical leakage to be $\lesssim 0.1\%$ by two methods. 
Its polarization angle was consistent with expectation.
The total leakage, measured from two different types of data, was found to be larger than the optical leakage as $\sim0.4\%\,(0.1\%)$ for real (imaginary) part, 
which indicates that the non-linearities of the detector have significant contributions.
The ratio of the total leakage into real and imaginary parts was in good agreement with the expectation from the physical model of the \ac{TES}.
Although the amplitudes of measured total leakage were larger than our expectation from the simple TES model that may underestimate the transition shape dependent parameter,
none of the other possible sources of leakage we have considered can explain the large difference between the optical leakage and total leakage.
Our circumstantial evidence thus supports the idea that the detector non-linearity is the main source of the total leakage.

We also performed template subtraction using our own intensity timestream,
and have found that we can efficiently remove the leakage and improve the 1/f noise by an order of magnitude in \acp{PSD}. 
The 1/f knee frequency for the polarization timestream averaged across the focal plane was $32\,(15)\,\mathrm{mHz}$ for real (imaginary) part, which roughly corresponds to multipole $\ell \simeq 39\,(18)$.
The systematic uncertainty for B-mode power spectrum measurement due to the residual \ac{I2P} leakage is at most the level of $r \sim 0.03$, which can be suppressed by sky rotation and averaging among observations by orders of magnitudes, thus we would be able to mitigate the error to the level of $r<0.01$. 
These results are sufficient for the current level of sensitivity.

Future CMB experiments are expected to sharply increase their sensitivity, with a concomitant decrease in the allowed systematics budget. 
Further mitigations of the \ac{I2P} leakage and residual 1/f noise will be essential to the success of these experiments.
Possible approaches for further work include improving the temperature stability of the mirrors, focal plane, and readout electronics, and reducing detector non-linearity by operating the TES detectors at a higher loop-gain and/or in a resistance-locked loop.

\acknowledgments
S. Takakura was supported by Grant-in-Aid for JSPS Research Fellow.
The \pb project is funded by the National Science Foundation under Grants No. AST- 0618398 and No. AST-1212230.
The James Ax Observatory operates in the Parque Astron\'omico Atacama in Northern Chile under the auspices of the Comisi\'on Nacional de Investigaci\'on Cient\'ifica y Tecnol\'ogica de Chile (CONICYT). 
This research used resources of the Central Computing System, owned and operated by the Computing Research Center at KEK,
the HPCI system (Project ID:hp150132), and the National Energy Research Scientific Computing Center, a DOE Office of Science User Facility supported by the Office of Science of the U.S. Department of Energy under Contract No. DE-AC02-05CH11231.
In Japan, this work was supported by MEXT KAKENHI Grant Numbers JP15H05891, 21111002, JSPS KAKENHI Grant Numbers JP26220709, JP24111715, and JSPS Core-to-Core Program, A. Advanced Research Networks.
In Italy, this work was supported by the RADIOFOREGROUNDS grant of the European Union's Horizon 2020 research and innovation programme (COMPET-05-2015, grant agreement number 687312) as well as by the INDARK INFN Initiative.
MA acknowledges support from CONICYT's UC Berkeley-Chile Seed Grant (CLAS fund) Number 77047, Fondecyt project 1130777, DFI postgraduate scholarship program and DFI Postgraduate Competitive Fund for Support in the Attendance to Scientific Events.
DB acknowledges support from NSF grant AST-1501422.
GF acknowledges the support of the CNES postdoctoral program.
JP acknowledges support from the Science and Technology Facilities Council [grant number ST/L000652/1] and from the European Research Council under the European Union's Seventh Framework Programme (FP/2007-2013) / ERC Grant Agreement No. [616170].
CR acknowledges support from an Australian Research Council's Future Fellowship (FT150100074).

\bibliography{reference}
\end{document}